\documentclass[fleqn,usenatbib]{mnras}
\usepackage{newtxtext,newtxmath}

\usepackage[T1]{fontenc}

\DeclareRobustCommand{\VAN}[3]{#2}
\let\VANthebibliography\thebibliography
\def\thebibliography{\DeclareRobustCommand{\VAN}[3]{##3}\VANthebibliography}

\usepackage{graphicx}	% Including figure files
\usepackage{amsmath}	% Advanced maths commands
\usepackage{pdflscape}
\usepackage{subcaption}
\usepackage{lineno}
\usepackage{bm}

\newcommand{\rbracket}{]}

\title[$T_e$-metallicities of galaxies over cosmic time]{The \textit{JWST} EXCELS survey: Probing strong-line diagnostics and the chemical evolution of galaxies over cosmic time using $T_e$-metallicities}

\author[D. Scholte et al.]{
D. Scholte,$^{1}$\thanks{E-mail: dscholte@ed.ac.uk (DS)}
F. Cullen,$^{1}$
A. C. Carnall,$^{1}$
K. Z. Arellano-C\'ordova,$^{1}$
T. M. Stanton,$^{1}$
L. Barrufet,$^{1}$
\newauthor{
C. T. Donnan,$^{1}$
J. S. Dunlop,$^{1}$
H.-H. Leung,$^{1}$
D. J. McLeod,$^{1}$
R. J. McLure,$^{1}$
J. M. Moustakas,$^{2}$
}
\newauthor{
C. L. Pollock,$^{3}$
A. E. Shapley,$^{4}$
S. Stevenson$^{1}$
and H. Zou$^{5,6}$
}
\\
$^{1}$Institute for Astronomy, University of Edinburgh, Royal Observatory, Edinburgh, EH9 3HJ, UK\\
$^{2}$Department of Physics and Astronomy, Siena College, 515 Loudon Road, Loudonville, NY 12110, USA\\
$^{3}$The Cosmic Dawn Center, Niels Bohr Institute, University of Copenhagen, Jagtvej 128, DK-2200 Copenhagen N, Denmark\\
$^{4}$Department of Physics \& Astronomy, University of California, 430 Portola Plaza, Los Angeles CA 90095, USA\\
$^{5}$Key Laboratory of Optical Astronomy, National Astronomical Observatories, Chinese Academy of Sciences, Beijing 100101, People's Republic of China\\
$^{6}$School of Astronomy and Space Science, University of Chinese Academy of Sciences, Beijing 101408, People's Republic of China\\
}

\date{Accepted XXX. Received YYY; in original form ZZZ}

\pubyear{\the\year{}}

\begin{document}
\label{firstpage}
\pagerange{\pageref{firstpage}--\pageref{lastpage}}
\maketitle

% Abstract of the paper
\begin{abstract}
We present an analysis of the rest-frame optical spectra of 22 [\textsc{Oiii}]$\lambda$4363 detected galaxies in the redshift range $1.65 \leq z \leq 7.92$ (with $\langle z\rangle = 4.05$) from \emph{JWST}/NIRSpec medium-resolution observations taken as part of the EXCELS survey.
To supplement these high-redshift sources, we also consider a sample of 782 local [\textsc{Oiii}]$\lambda$4363 detected galaxies from the DESI Early Data Release. 
Our analysis demonstrates that many strong-line calibrations are biased in the early Universe due to the systematic evolution in ionization conditions with redshift.
However, the recently introduced $\widehat{R}$ calibration mostly removes the dependence on ionization state and can be considered a largely redshift-independent calibration.
In a similar spirit, we introduce a new strong-line diagnostic, $\widehat{RNe}$, which can be used to robustly estimate metallicities when the [\textsc{Oiii}]$\lambda$5007 is redshifted out of the wavelength range of \emph{JWST}/NIRSpec at $z > 9.5$.
We also show that strong-line diagnostics using the [\textsc{Nii}]$\lambda$6584 emission line are likely to be biased at high-redshift due to a moderate enhancement in the average N/O abundance ratios (at fixed O/H) in these sources.
Finally, we discuss the location of our new [\textsc{Oiii}]$\lambda$4363 detected galaxies at $z\simeq4$ on the mass-metallicity plane and investigate the redshift evolution of the fundamental metallicity relation (FMR). 
We find tentative evidence for an increasing deviation from the FMR at $z > 4$ which might indicate fundamental differences in the baryon cycle at these redshifts.
However, more data are required as our high-redshift constraints are still based on a relatively small sample of galaxies and the significance of the deviation is strongly dependent on the assumed form of the fundamental metallicity relation.
\end{abstract}


\section{Introduction} \label{sec:intro}
The metal enrichment of the interstellar medium (ISM) is an important probe for the chemical evolution of galaxies over cosmic time. The abundances of metals in the gas phase of galaxies are the result of the interplay of processes such as inflows of pristine gas, metal production in stars and outflows of metal-enriched gas \citep[e.g.,][]{finlator2008}. The relative abundances of different elements also provide information on the type of stars responsible for their production. The production and release of the elements occurs through several pathways in different types of stars and on different time scales \citep[for a review, see][]{kobayashi2020}, therefore multi-element abundance ratios also probe the star formation histories of galaxies.

The oxygen abundance is commonly used as a proxy for metallicity because it is generally the most abundant metal and readily observed through emission lines in the rest-frame optical spectrum of star-forming galaxies and individual H\textsc{ii}-regions. We will refer to oxygen abundance as metallicity throughout the rest of this work. There are several methods to infer the gas-phase metallicity of galaxies; the main methods are (1) the recombination line method \citep[e.g.,][]{esteban2009, esteban2014}, (2) the electron temperature ($T_e$) sensitive method \citep[e.g.,][]{peimbert1967, stasinska1982, campbell1986, garnett1992}, (3) photoionisation modelling \citep[e.g.,][]{ferland1996, mcgaugh1991, charlot2001, kewley2002, byler2017, ferland2017} and (4) strong-line calibrations \citep[e.g.,][]{pagel1979, alloin1979, storchi-bergmann1994, pettini2004, curti2020}. There are systematic discrepancies between the metallicities derived using these different methods. The absolute abundance scale of galaxies is therefore debated, with discrepancies up to $\sim 0.7$ dex \citep[][]{moustakas2010,stasinska2010,kewley2019review}. These discrepancies can have a significant influence on inferred galaxy scaling relations as shown by \cite{kewley2008}. Nonetheless, the relative abundances of galaxies derived using the same method are typically regarded as robust. In this work, we mainly focus on the $T_e$-method and strong-line calibrations derived using samples of galaxies with electron temperature measurements. 

The $T_e$-method is regarded as one of the most reliable metallicity measurements. This technique uses auroral emission lines such as [\textsc{Oiii}]4363, [\textsc{Nii}]$\lambda$5755, [\textsc{Sii}]$\lambda$6312 or the [\textsc{Oii}]$\lambda\lambda$7320,7330 doublet which can be used to directly constrain the electron temperature of the ionised gas in \textsc{Hii}-regions purely based on atomic properties and collisional excitation calculations \citep[e.g.,][]{aller1984}. The auroral emission lines are typically faint and therefore only observable for particular populations of metal-poor, highly star-forming galaxies \citep[e.g.,][]{izotov2006, zou2024}, using very deep spectroscopic observations \citep[e.g.,][]{esteban2004, strom2023}, or through stacking of the spectra of many galaxies \citep[e.g.,][]{andrews2013, curti2017}. Due to the challenges in observing these auroral emission lines, the samples of high-redshift galaxies with $T_e$-metallicities are still very small. However, deep \textit{JWST}/NIRSpec observations are slowly building up a sample of auroral detections \citep[e.g.][]{arellano-cordova2022,curti2023, nakajima2023, morishita2024, welch2024}. In this work, we add a significant number of new detections.

There have been extensive efforts to infer the metallicity of galaxies solely based on strong/bright emission lines of galaxies. However, the emission line strengths of the strong emission lines are dependent on multiple physical properties of the irradiated ISM. Many strong-line diagnostics are calibrated using limited samples of mostly local galaxies with $T_e$-metallicity measurements \citep[e.g.,][]{denicolo2002, pettini2004, curti2017, bian2018, sanders2021, nakajima2022} and recently some small samples of high-redshift galaxies \citep[e.g.,][]{sanders2024, chakraborty2024}. Therefore, it is important to verify that locally calibrated strong-line methods can be applied to the measurement of metallicities in the high-redshift Universe, where there are systematic differences in the ISM conditions of the galaxy population, such as enhanced alpha-element abundances, increased ionisation parameter at fixed metallicity, harder ionising radiation and likely systematically enhanced nitrogen-to-oxygen abundance ratios at fixed metallicity \citep{steidel2014, shapley2015, sanders2016, kashino2017, strom2017, strom2018, topping2020, cullen2021, kashino2022, sanders2023excitation, shapley2024, chartab2024, stanton2024kmos, arellano-cordova2024excels, stanton2024excels}. Recent \textit{JWST}/NIRSpec \citep[][]{jakobsen2022, ferruit2022, gordon2022} observations have already resulted in the detection of auroral emission lines in tens of high-redshift galaxies. Many local strong-line calibrations do not perform well at high redshift due to the systematic evolution of the ionisation parameter for star-forming regions in early galaxies \citep[as shown by e.g.,][]{sanders2024}. The bias introduced by this evolution can be addressed by either accounting for the dependence in the strong-line calibration \citep[e.g., the $\rm EW(H\beta)$ dependent calibrations by][]{nakajima2022} or by constructing a strong-line calibration that is insensitive to the ionisation parameter \citep[e.g., using the $\widehat{R}$ line ratio introduced by][]{laseter2024}.

In the nearby Universe and increasingly at high redshift, the relation between the gas-phase metallicity and other properties of galaxies has been mapped in great detail. In particular, there is a strong correlation with stellar mass, which is referred to as the mass-metallicity relation \citep[MZR;][]{lequeux1979,tremonti2004, gallazzi2005, andrews2013, curti2020, scholte2024}. The metallicity of galaxies steadily increases with stellar mass, which is typically explained through the decreasing fraction of metals that are expelled from the ISM through outflows as a result of the deeper potential wells galaxies reside in at higher mass \citep{recchi2008, dave2012, lilly2013}. The evolution of the MZR over redshift has also been studied extensively. Until the pre-\textit{JWST} era, this had been reliably mapped up to $z \sim 3.5$ \citep[for a review, see][]{maiolino2019review}. The MZR is systematically shifted to lower metallicities at higher redshift \citep[e.g.,][]{savaglio2005,moustakas2011,zahid2013,cullen2014,sanders2021, cullen2019, kashino2022, chartab2024, stanton2024kmos}. The redshift evolution of the MZR can to some degree be understood in terms of the correlation between stellar mass, metallicity and star formation rate \citep[e.g.,][]{ellison2008}. It has also been shown that galaxies reside on a redshift invariant 2D plane in the three-parameter space spanned by stellar mass, metallicity and star formation rate; this is referred to as the fundamental metallicity relation \citep[FMR; e.g.,][]{mannucci2010, lara-lopez2010, andrews2013, curti2020}. This correlation may be even stronger as a function of the gas content of galaxies which is, however, more challenging to constrain \citep[e.g.,][]{brinchmann2013, bothwell2013, bothwell2016, lara-lopez2013, hughes2013,brown2018, scholte2023, scholte2024}. 

The advent of \textit{JWST} has made it possible to push measurements of the MZR to much higher redshift. This is a result of both the infrared coverage of the telescope and its increased sensitivity compared to previous generations of telescopes. Therefore, in recent years, studies of the MZR have been pushed to both higher redshift and low-mass galaxies \citep[e.g.,][]{arellano-cordova2022,li2023, curti2023,nakajima2023}. Some of these studies have found that at high-redshift ($z>4$) galaxies become systematically offset from the FMR \citep[e.g.,][]{heintz2023, curti2024}, which could be a tantalising indication that we may be probing the era of galaxy formation where a balance between gas accretion, chemical enrichment and feedback processes has not yet been reached \citep{tacconi2020review}. These exciting results are drawing increased scrutiny onto the use of strong-line metallicity measurements. Luckily, the growing number of $T_e$-metallicity measurements at these redshifts are starting to make it possible to verify these results \citep[e.g.,][]{arellano-cordova2022,sanders2023metallicity, curti2023,nakajima2023, laseter2024, morishita2024}.

In this work we present our measurements of $T_e$-metallicities and other properties of galaxies observed with \textit{JWST}/NIRSpec in the \textit{JWST} EXCELS survey \citep[][]{carnall2024}. The EXCELS survey provides some of the deepest, medium-resolution ($R\sim1000$) \textit{JWST} spectroscopy of high-redshift galaxies to date. We present the measurements of a total of 22 galaxies with detections of the [\textsc{Oiii}]$\lambda$4363 emission line. These include measurements of the galaxies discussed in \cite{stanton2024excels} and \cite{arellano-cordova2024excels}. We supplement these high-redshift observations with a new sample of local $T_e$-metallicity measurement using data from the Dark Energy Survey Early Data Release \citep[DESI EDR;][]{desi2022, desi2023edr}. Due to the increased depth of the DESI Survey compared to previous generations of large spectroscopic surveys \citep[e.g., SDSS;][]{york2000} this local sample contains many faint and low-mass galaxies which previously had not been observed spectroscopically. We use these combined samples to investigate how successful various strong-line diagnostics are at retrieving the metallicity of galaxies in the early Universe.

In Section \ref{sec:observations} we discuss the details of the observations that are used in this work. The data reduction methods can be found in Section \ref{sec:data}. Our further analysis methods are described in Section \ref{sec:derived_data}. In the following sections we show our results on the general properties of our \textit{JWST} EXCELS sample in Section \ref{sec:properties}, comparisons of strong-line calibrations in Section \ref{sec:strong_lines} and measurements of the mass-metallicity relation/fundamental metallicity relation in Section \ref{sec:mzr}. Finally, we discuss these results in Section \ref{sec:discussion} and share our conclusions and future outlook in Section \ref{sec:conclusions}.

We repeatedly use several emission line ratios for which we define the names below:
\begin{align*}
    R2 &= \log \left[ \frac{[\textsc{Oii}]\lambda\lambda3726,3729}{\textsc{H}\beta} \right], \\
    R3 &= \log \left[ \frac{[\textsc{Oiii}]\lambda5007}{\textsc{H}\beta} \right], \\
    O32 &= \log \left[ \frac{[\textsc{Oiii}]\lambda\lambda4959,5007}{[\textsc{Oii}]\lambda\lambda3726,3729} \right], \\
    R23 &= \log \left[ \frac{[\textsc{Oii}]\lambda\lambda3726,3729+[\textsc{Oiii}]\lambda\lambda4959,5007}{\textsc{H}\beta} \right], \\
    \widehat{R} &= 0.47 \times R2 + 0.88 \times R3, \\
    N2 &= \log \left[ \frac{[\textsc{Nii}]\lambda6584}{\textsc{H}\alpha} \right], \\
    O3N2 &= \log \left[ \frac{[\textsc{Oiii}]\lambda5007/\textsc{H}\beta}{[\textsc{Nii}]\lambda6584/\textsc{H}\alpha} \right], \\
    N2O2 &= \log \left[ \frac{[\textsc{Nii}]\lambda6584}{[\textsc{Oii}]\lambda\lambda3726,3729} \right], \\
    S2 &= \log \left[ \frac{[\textsc{Sii}]\lambda\lambda6716,6731}{H\alpha} \right], \\
    Ne3 &= \log \left[ \frac{[\textrm{Ne}\textsc{iii}]\lambda3869}{\textsc{H}\beta} \right], \\
    Ne3O2 &= \log \left[ \frac{[\textrm{Ne}\textsc{iii}]\lambda3869}{[\textsc{Oii}]\lambda\lambda3726,3729} \right], \\
    \widehat{RNe} &= 0.47 \times \log \left[ \frac{[\textsc{Oii}]\lambda\lambda3726,3729}{\textsc{H}\gamma} \right]\\ 
     & \quad + 0.88 \times \log \left[ \frac{[\textrm{Ne}\textsc{iii}]\lambda3869}{\textsc{H}\gamma} \right]. \\
\end{align*}
Before the calculation of the line ratio each of these emission lines are dust corrected. We assume a standard cosmological model with $H_0=70$\,km s$^{-1}$ Mpc$^{-1}$, $\Omega_m=0.3$ and $\Omega_{\Lambda}=0.7$ and a solar oxygen abundance of $\mathrm{12+log(O/H)} = 8.69$ \citep{asplund_2021_solar_abn}.

\section{Observations and data products} \label{sec:observations}

\subsection{EXCELS survey}
Our main sample of observations is drawn from the galaxies observed as part of the \textit{JWST} EXCELS spectroscopic survey \citep[GO 3543; PIs: Carnall, Cullen;][]{carnall2024}. This \textit{JWST}/NIRSpec survey provides deep, medium-resolution ($R\sim1000$) spectroscopy for galaxies in two general categories: massive quiescent galaxies below $z \sim 5$ and star-forming galaxies below $z \sim 8$. In this work we focus on the second category of star-forming galaxies; they are selected primarily from the VANDELS survey catalogue \citep[][]{galametz2013,mclure2018} and supplemented by the \textit{JWST} Cycle 1 PRIMER UDS imaging \citep[][]{dunlop2021,mcleod2024} at $z \geq 5$. The targeted galaxies were observed using the G140M/F100LP, G235M/F170LP and G395M/F290LP grating/filter combinations, providing coverage of the rest-frame UV, optical and sometimes infrared spectra. Targets were observed for $\simeq 4$ hrs in G140M and G395M and $\simeq 5$ hrs in G235M; not all targets were observed in all three gratings. A detailed summary of the sample selection and survey strategy can be found in \cite{carnall2024}.

\subsection{DESI Early Data Release} \label{sec:desi_data}
To supplement this new high-redshift sample, we use data from the DESI Early Data Release as a low-redshift supplement to our high-redshift spectra from the EXCELS programme. The Dark Energy Spectroscopic Instrument (DESI) is a spectroscopic instrument on the Mayall 4-meter telescope at Kitt Peak National Observatory \citep{desi2022, desi2023}. The DESI Early Data Release (EDR) contains observations taken during the five-month Survey Validation (SV) period \citep[][]{desi2023, desi2023edr}. The observations of galaxies in the Bright Galaxy Survey (BGS), the low redshift portion of the Emission Line Galaxies (ELG) target class and targets observed as part of the LOW-Z secondary target programme are of particular interest for this work as these programmes observe the majority of DESI low-redshift targets and in particular low-metallicity dwarf galaxies \citep{hahn2023bgs, darragh-ford2023}. The Bright Galaxy Survey is a two-tier $r-$band magnitude limited survey; the survey is split into BGS Bright with $r < 19.5$ and BGS Faint with $19.5 < r < 20.175$. The LOW-Z secondary target programme is designed to observe a sample of faint low mass dwarfs with $r<21.0$. A significant fraction of the observations taken in the first period of survey validation (SV1) are significantly deeper than observations in the remainder of survey validation or DESI main survey observations. The exposure times during most of this period were a factor of four deeper than main survey observing. Therefore, the SV1 observations are particularly useful in this work where we require the detection of the faint [\textsc{Oiii}]$\lambda$4363 auroral oxygen emission line. The emission line measurements for the DESI EDR sample are taken from the \textsc{FastSpecFit} value added catalogue\footnote{https://data.desi.lbl.gov/doc/releases/edr/vac/fastspecfit/} \citep{moustakas2023}. \textsc{FastSpecFit} is a stellar continuum and emission line modelling code optimised for DESI, which provides emission line measurements for all DESI galaxies. The stellar mass and star formation rate measurements for DESI galaxies are taken from the DESI stellar mass catalogue\footnote{https://data.desi.lbl.gov/doc/releases/edr/vac/stellar-mass-emline/} generated using Cigale \citep[][]{boquien2019} SED fitting to the DESI Legacy Imaging Surveys photometry \citep[][]{dey2019} and spectroscopy \citep[for a detailed description see][]{zou2024}.

\section{Data reduction and sample selection} \label{sec:data}

\subsection{EXCELS data reduction}
We initially process the raw level 1 data products from the Mikulski Archive for Space Telescopes (MAST) using v1.15.1 of the \textit{JWST} reduction pipeline\footnote{https://github.com/spacetelescope/jwst}. We use the default level 1 configuration with the inclusion of advanced snowball rejection and the CRDS\_CTX = jwst\_1258.pmap version of the \textit{JWST} Calibration Reference Data System (CRDS) files. Following this step, we apply the level 2 pipeline in default configuration. Finally, the level 3 pipeline provides the final combined 2D spectra that feed into our analysis. The 1D spectra are extracted using a custom extraction method from the 2D spectra \citep{horne1986}, where we set the extraction centroid as the flux-weighted mean position of the object within the NIRSpec/MSA slitlet \citep[for details see][]{carnall2024}.

A two-step flux calibration is needed to derive the absolute flux calibrated spectra. First we perform relative flux calibration of the spectra in each of the gratings. Then we apply a correction for the wavelength dependent slit losses due to the increasing FWHM of the PSF with wavelength. We also need to correct for small offsets of the source position within the slit which can introduce mismatches in the flux levels between gratings. We scale the fluxes of the different gratings using the overlapping wavelength regions of the gratings. The G140M and/or G395M spectra are scaled to the G235M spectra. After this relative flux calibration we perform an absolute flux calibration where we calibrate the spectra to the broadband photometry \citep[using a similar approach to the CECILIA survey;][]{rogers2024}. We use the deep HST, \textit{JWST}/NIRCam \citep{dunlop2021,mcleod2024} and \textit{JWST}/MIRI imaging available for the galaxies observed in EXCELS (HST filters F435W, F606W and F814W; \textit{JWST}/NIRCam filters F090W, F115W, F150W, F200W, F277W, F356W, F444W, F410M; \textit{JWST}/MIRI filter F770W). The HSC and \textit{JWST}/NIRCam broadband fluxes that are extracted using a 0.5 arcsec diameter aperture from F444W PSF-homogenised images and corrected to total flux using the F356W FLUX\_AUTO measurements from \textsc{SourceExtractor} \citep{bertin1996}. Total flux was measured in the \textit{JWST}/MIRI F770W filter using 0.7 arcsec diameter apertures. Each spectrum is integrated through the overlapping filters and then scaled to the imaging fluxes using a linearly interpolated wavelength dependent correction between bands. The reliability of these flux corrections are verified by observed emission lines in wavelength ranges where the gratings overlap; the scatter in the flux measurements of the same emission lines in multiple gratings is $\sim$8 per cent.

\begin{figure*}
	\begin{center}
	\includegraphics[width=0.85\textwidth]{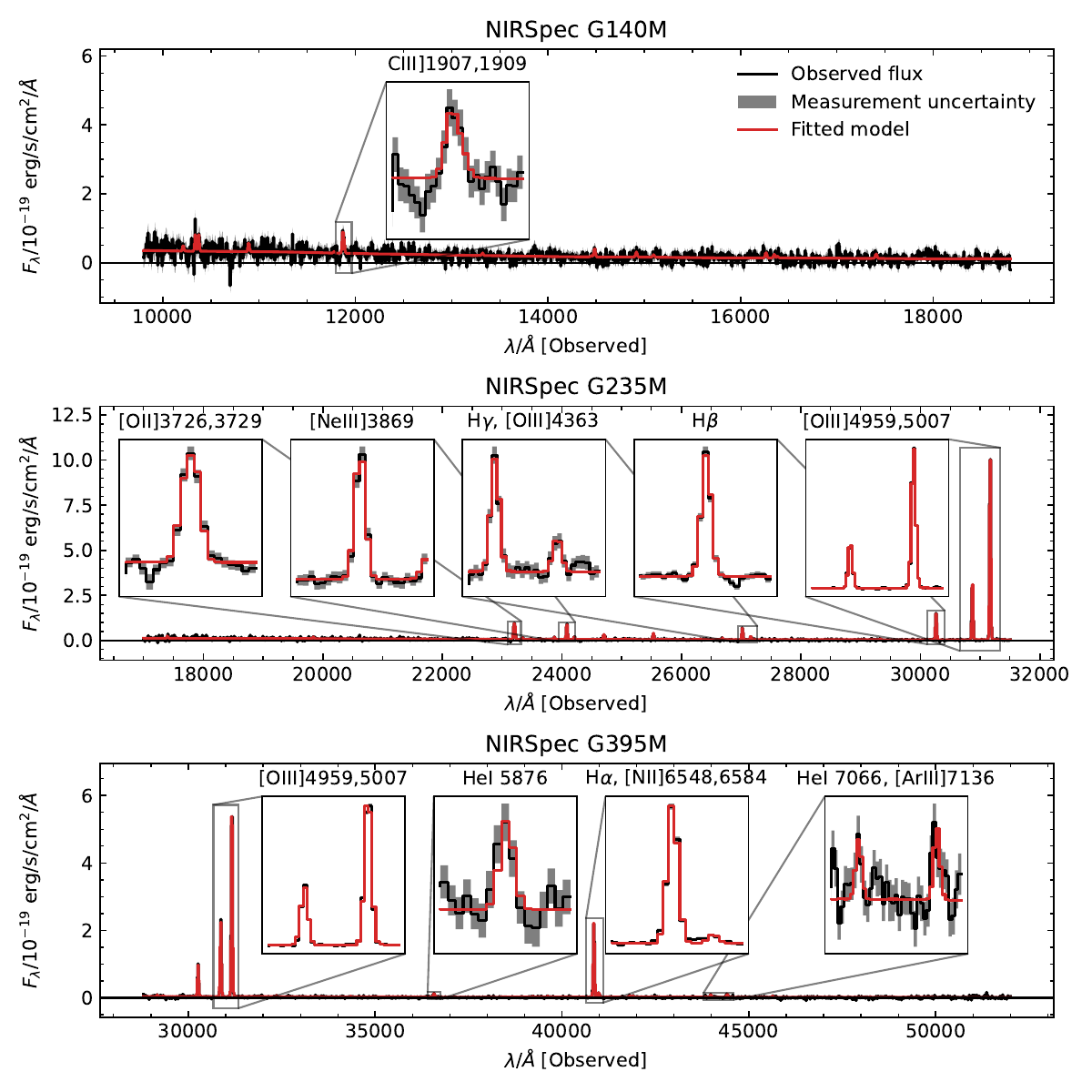}
	\caption{An example of a \textit{JWST} EXCELS spectrum (\textit{black, grey 1$\sigma$ uncertainties}) of EXCELS-121806 \citep[see also][]{arellano-cordova2024excels} shown together with the fitted continuum and emission line measurements (\textit{red}). The three panels show the observed flux in the NIRSpec G140M (\textit{top}), G235M (\textit{middle}), G395M (\textit{bottom}) gratings. Inset panels zoom in on the strongest emission lines observed in each grating. Due to overlap in the wavelength range of the gratings some emission lines are observed in multiple gratings (e.g., [\textsc{Oiii}]$\lambda$5007).}
	\label{fig:emline_fit_example}
	\end{center}
\end{figure*}

\subsection{EXCELS emission line measurements}
We measure the emission line fluxes and equivalent widths of the reduced and calibrated 1D EXCELS spectra using a simple continuum subtraction and emission-line fitting routine. The continuum flux, $F_{\rm cont}$, is estimated using a running mean of the 16$^{\rm th}$ to 84$^{\rm th}$ percentile flux values within a top hat function with a rest-wavelength width of 350 \AA,
\begin{equation}
    F_{{\rm cont},i} = \left(\sum_{t}^n F_{t}\right) / n, ~ \textrm{where}~F_{16} \leq F_t \leq F_{84}
\end{equation}
where $F_t$ is the observed flux in pixel $t$ within the top hat and $n$ is the number of pixels where the flux values within the top hat window are between $F_{16}$ and $F_{84}$ \citep[][]{sdss_edr2002}. The choice to only include the flux densities within the 16$^{\rm th}$ to 84$^{\rm th}$ percentile range effectively masks the contamination of the continuum measurement by strong emission and absorption features. We estimate the standard deviation and standard error of the continuum flux measurement within the top hat window as follows
\begin{equation}
    {\rm STD}_{{\rm cont},i}\approx \frac{F_{84}-F_{16}}{2},
\end{equation}

\begin{equation}
    {\rm SE}_{{\rm cont},i} = \frac{{\rm STD}_{{\rm cont},i}}{\sqrt{n}}.
\end{equation}

Before the emission line measurement we also re-estimate the pixel flux measurement uncertainties by introducing a correction factor based on the residuals after continuum subtraction. The uncertainties derived from version v1.15.1 of the data reduction pipeline are typically underestimates of the true flux uncertainties in the wavelength bins. We implement the correction as follows 
\begin{equation}
    \sigma_{F} = \frac{\frac{1}{2}(R_{84}-R_{16})}{\tilde{\sigma}_{F,\rm pipe}} \times \sigma_{F,\rm pipe},
\end{equation}
where $R_{16}$, $R_{84}$ are the 16$^{\rm th}$ and 84$^{\rm th}$ percentiles of the continuum subtracted residuals ($R$), $\tilde{\sigma}_{F,\rm pipe}$ is the median of the pipeline flux uncertainty, $\sigma_{F,\rm pipe}$, and $\sigma_{F}$ is the corrected flux uncertainty used throughout this work. The median correction factor is $\sigma_{F} \sim1.6~\sigma_{F,\rm pipe}$. This is consistent with the correction factors derived in independent analyses \citep[][]{maseda2023, carnall2024}.

We measure the flux of the emission lines in the spectra using Gaussian line fits. All lines are fitted simultaneously to the continuum subtracted flux. The total emission line model $M(\lambda)$ is given by
\begin{equation}
    M(\lambda) = \sum_l A_l \times \exp\left[-\frac{1}{2} \frac{\left(\log{\lambda} - \log{\lambda_{l}^{\rm obs}}\right)^{2}}{\sigma_{\rm tot}^{2}}\right] ,
\end{equation}
where $A_l$ and $\log \lambda_{l}^{\rm obs}$ are the amplitude and logarithm of the observed wavelength of the emission line $l$. The observed wavelength of an emission line is given by 
\begin{equation}
    \lambda_{l}^{\rm obs} = \lambda_{l}^{\rm rest} \times \left(1+z + \frac{V_{l}}{c} \right),
\end{equation}
where $\lambda_{l}^{\rm rest}$ is the rest-wavelength of the emission line, $c$ the speed of light, $z$ the redshift of the galaxy and $V_{l}$ the velocity of the emission line. The total width of an emission line, $\sigma_{\rm tot}$, is a convolution of intrinsic broadening and instrumental broadening dependent on the line wavelength and the NIRSpec grating used: $\sigma_{\rm tot}^2 = \sigma_{\rm intrinsic}^2 + \sigma_{\rm grating}^2(\lambda)$. The emission line width is given in units of log \AA{} which can be converted to conventional units of $\sigma[\rm km~s^{-1}] = \sigma[\rm log ~$\AA$] \times c[\rm km~s^{-1}] / ln(10)$. The emission lines are fitted using a common line velocity and intrinsic line width. The amplitude of the emission line is a free parameter for each individual line included in the fitting; this results in $2 + n_{\rm lines}$ free parameters. The emission line fitting is performed using weighted least-squares minimisation: $ \chi^2 = \sum_i \frac{(R_i - M_i)^2}{\sigma_{F,i}^2}$, with $i$ the wavelength bins in the observed spectrum. The $\chi^2$-minimisation is performed with the \textsc{SciPy} least squares implementation \citep{branch1999, scipy2020}.

To measure the line fluxes and associated uncertainties we use a similar approach as used for the DESI EDR emission line catalogue \citep[\textsc{FastSpecFit;}][]{moustakas2023}. The emission line flux and uncertainty are calculated using line profile weighted measurements, where $P_i = M_i/\sum_i G_i$ is the normalised profile of a model emission line ($G_i$). The integrated Gaussian line flux is expressed as
\begin{equation}
    F_{\rm line} = \frac{\sum_i w_i ~\Delta \lambda_i ~G_i/P_i}{\sum_i w_i},
\end{equation}
where $\Delta \lambda_i$ is the observed-frame pixel width, and the weights $w_i$ are defined as
\begin{equation}
    w_i = \frac{P_i}{\left(\Delta \lambda ~\sigma_{F,i}\right)^2}.
\end{equation}
The uncertainty on the emission line flux is given by
\begin{equation}
    \sigma_{\rm line} = \sum w_i^{-\frac{1}{2}}.
\end{equation}
The line flux measurements are integrated within a 3$\sigma$ region around the centre of each emission line. The equivalent width ($F_{\rm line}/F_{\rm cont}(\lambda_{\rm line})$) uncertainty is calculated using error propagation of the uncertainties on the line and continuum flux measurements. 

These uncertainties do not include any contribution from the flux calibration process. Therefore, an additional uncertainty contribution is included when absolute flux values or flux ratios of emission lines in multiple gratings are used. We estimated this uncertainty factor using the emission lines that are observed in the overlapping wavelength coverage of multiple gratings. This yields an average uncertainty due to flux calibration of $\sim$8 per cent of the emission line flux.

Lastly, the measurement of the Balmer emission lines is affected by Balmer absorption due to stellar absorption lines in the continuum flux. We correct for this Balmer absorption using the Balmer absorption fluxes derived from Lorentzian line fits to the model spectra produced using \textsc{Bagpipes}. A more detailed overview of the SED fitting to the photometry can be found in Section \ref{sec:bagpipes}. We apply a correction using an emission filling fraction of 30 per cent of the Balmer absorption flux for each of the Balmer emission lines. This is similar to the emission filling fractions reported by \cite{reddy2018}. The median Balmer absorption correction for our EXCELS galaxies is 0.12, 1.4, 4.2 per cent on the H$\alpha$, H$\beta$, H$\gamma$ emission lines, respectively.

\subsection{EXCELS sample selection}
The sample selection of the EXCELS survey galaxies in this paper is driven by the need to detect the faint [\textsc{Oiii}]$\lambda$4363 auroral emission line. To ensure robust detections we only include [\textsc{Oiii}]$\lambda$4363 $\rm S/N \geq 3$. For our analysis, we also require a set of strong emission lines; these are typically much brighter than the [\textsc{Oiii}]$\lambda$4363 emission line. However, we apply a $\rm S/N \geq 2$ selection on the flux of the [\textsc{Oii}]$\lambda\lambda$3726,3729 doublet, H$\beta$, [\textsc{Oiii}]$\lambda$4959, [\textsc{Oiii}]$\lambda$5007 and either H$\alpha$ or H$\gamma$. The requirement of a detection of either H$\alpha$ or H$\gamma$ allows for the inclusion of galaxies where H$\alpha$ is redshifted out of the observable wavelength range of the NIRSpec gratings while maintaining a constraint on the dust attenuation. These selections yield a total sample of 22 galaxies. Our EXCELS sample covers a redshift range between $1.65<z<7.92$ with a median redshift $\langle z \rangle = 4.05$. We find no evidence of AGN activity in these galaxies using the [\textsc{Nii}]-BPT diagram \citep{baldwin1981} where available and the [\textsc{Oiii}]$\lambda$5007/[\textsc{Oiii}]$\lambda$4363 versus [\textsc{Oiii}]$\lambda$4363/H$\gamma$ diagram \citep[][]{mazzolari2024}. We also do not find broad emission lines or abnormal measurements in $\rm 12+log(O/H)$ versus $T_e$ space. We note that EXCELS-93897 has a high log([\textsc{Oiii}]$\lambda$5007/H$\beta$)$~=1.03 \pm 0.04$ ratio, however, no convincing indications of AGN activity.

\subsection{DESI sample selection}
The choices made in the sample selections of the DESI EDR data are based on the motivation to provide a high-fidelity local comparison sample to our high-redshift measurements. Therefore, the selection criteria for this dataset are comparatively strict. To ensure the selection of high-quality galaxy spectra with successful redshift measurements from the DESI EDR catalogues we impose the following selection criteria based on the DESI data reduction pipeline outputs\footnote{\url{https://desidatamodel.readthedocs.io/}}: \texttt{SPECTYPE == GALAXY}, \texttt{ZWARN == 0} and \texttt{DELTACHI2 >= 40}. As we require the detection of strong emission line ratios in this work we impose a $\rm S/N>3$ threshold for the line flux of [\textsc{Oii}]$\lambda\lambda$3726, 3729, H$\beta$, [\textsc{Oiii}]$\lambda\lambda$4959, 5007 and H$\alpha$. We further impose a high $\rm S/N > 5$ threshold on the [\textsc{Oiii}]$\lambda$4363 auroral emission line. This high threshold on the auroral emission line flux removes any spurious detections that may contaminate the sample in this work. This high threshold is to ensure we have a high-fidelity comparison sample. An important consideration here is that [\textsc{Oiii}]$\lambda$4363 emitting galaxies are much rarer in the DESI EDR sample than they are in EXCELS; this means that at lower S/N thresholds the contaminating fraction is significantly higher. We also require that \texttt{RCHI2\_LINE < 5} and that the \texttt{[LINE]\_FLUX} measurement derived using the Gaussian fitting does not deviate by more than 10 per cent from the \texttt{[LINE]\_BOXFLUX} measurement for the [\textsc{Oiii}]$\lambda$5007 and H$\alpha$ emission lines; this selection removes any spectra where the line fit failed. We only include spectra where $\rm EW(H\beta) > 20$; this ensures real emission lines are fitted rather than spurious fits to residuals in the subtracted continuum flux. We only include galaxies which are classified as star forming in the BPT-diagram by either the \cite{kewley2001} or the \cite{kauffmann2003} criteria.

\renewcommand{\arraystretch}{1.25}

\begin{table*}

\caption{The free parameters of the \textsc{Bagpipes} model we fit to our photometric data with their associated prior distributions.  The upper limit on $\tau$, $t_\mathrm{obs}$, is the age of the Universe as a function of redshift. Logarithmic priors are all in base ten.}
\begingroup
\setlength{\tabcolsep}{6pt}
\renewcommand{\arraystretch}{1.1}
\begin{tabular}{llllll}
\hline
Component & Parameter & Symbol / Unit & Range & Prior \\
\hline \hline
SFH & Total stellar mass formed & $M_*\ /\ \mathrm{M_\odot}$ & (1, $10^{13}$) & Logarithmic \\
& Stellar metallicity & $Z_*\ /\ \mathrm{Z_\odot}$ & (0.007, 2.45) & Logarithmic \\
& Double-power-law falling slope & $\alpha$ & (0.1, 1000) & Logarithmic \\
& Double-power-law rising slope & $\beta$ & (0.1, 1000) & Logarithmic \\
& Double-power-law turnover time & $\tau$ / Gyr & (0.1, $t_\mathrm{obs}$) & Uniform \\
\hline
Dust & $V-$band attenuation & $A_V$ / mag & (0, 4) & Uniform \\
\hline
\end{tabular}
\endgroup
\label{tab:bagpipes}
\end{table*}
\renewcommand{\arraystretch}{1.}

\subsection{Other literature data} \label{sec:literature_sample}
For some of our analysis we include a literature sample of galaxies which have been detected in other studies. We select this literature sample based on galaxies which have metallicity measurements derived using the $T_e$-method at $z>2$. We only include measurements of galaxies which also have reported stellar masses, star formation rates and metallicities as these measurements are required to study the fundamental metallicity relation. This sample includes 2 galaxies detected by \cite{sanders2023metallicity} in observations using the Keck/MOSFIRE instrument. It further includes detections in \textit{JWST} observations of 10 galaxies reported by \cite{nakajima2023} of which 3 galaxies were previously reported by \cite{curti2023}, 15 galaxies reported by \cite{morishita2024}, and the lowest metallicity EXCELS galaxy reported by \cite{cullen2025} which is not included in our sample due to the non-detection of [\textsc{Oii}]$\lambda\lambda$3726,3729, however, detailed modelling shows that this is due to a negligible abundance of singly ionised oxygen (O$^{+}$/H$^{+} \simeq0$).

\section{Derived data products} \label{sec:derived_data}
\subsection{\textsc{Bagpipes} SED fitting} \label{sec:bagpipes}
We use the \textsc{Bagpipes} code \citep{carnall2018, carnall2019} to model the spectral energy distribution (SED) of the galaxies in our EXCELS sample. We use photometric observations in 11 or 12 (depending on \textit{JWST}/MIRI 7.7um availability) photometric bands ranging from the rest-frame UV to IR to constrain the SEDs of these galaxies (HST filters F435W, F606W and F814W; \textit{JWST}/NIRCam filters F090W, F115W, F150W, F200W, F277W, F356W,F444W, F410M, \textit{JWST}/MIRI filter F770W). Photometry in the HST and \textit{JWST} NIRCam filters are available for all the galaxies in the dataset and the catalogue is described in \citet{mcleod2024} and \citet{cullen2024}; however, the \textit{JWST}/MIRI photometry is only available for roughly half the galaxies in our sample. We test the importance of including the MIRI photometry and find that there is no large impact on constraints on relevant parameters such as stellar mass or star formation rate derived without the inclusion of this photometric band. 

We subtract the contribution of emission lines from the total flux in each photometric band, as observed in the spectra. The most significant contributions are from the bright [\textsc{Oiii}]$\lambda\lambda$4959,5007 doublet and the H$\alpha$ emission line. Where there is no coverage of H$\alpha$ in the spectra we estimate the contribution to photometry using the H$\alpha$/H$\beta$ flux ratio. In the absence of a H$\alpha$ measurement we use the H$\beta$/H$\gamma$ ratio to determine the effect of dust extinction on the H$\beta$ flux. We then convert the unattenuated H$\beta$ flux to H$\alpha$ using an assumed intrinsic ratio of 2.86.

The details of the model configuration of \textsc{Bagpipes} fitting are shown in Table \ref{tab:bagpipes}. We use the \cite{bruzual2003} stellar population models and assume a \cite{chabrier2003} initial mass function. The star formation history (SFH) is modelled using a double power-law model. The dust is modelled using a \cite{salim2018} dust model where $A_{V}$ is a free parameter. The model constraints provide us with measurements of the stellar mass as well as average star formation rate estimates which are integrated over the last 10 Myr of the SFH. Our \textsc{Bagpipes} stellar mass and star formation rate measurements are listed in Table \ref{tab:excels_derived_data}.

\subsection{H$\alpha$ and H$\beta$ star formation rates}
We also calculate star formation rates from the H$\alpha$ or H$\beta$ flux, which is a more sensitive probe of the star formation rate on 10 Myr time scales than the broad band SED measurements. We use the \cite{reddy2018} calibration
\begin{equation}
    \frac{\dot{M}_{\star}}{\textrm{M}_{\odot}~\textrm{yr}^{-1}} = 3.236 \times 10^{-42} \left( \frac{L_{\rm H\alpha}}{\textrm{erg~s}^{-1}}\right),
\end{equation}
where $L_{\rm H\alpha}$ is the unattenuated H$\alpha$ luminosity. The luminosity to SFR conversion factor of this calibration is lower than a typical conversion factor of $4.634 \times 10^{-42}$ by e.g.,{} \cite{hao2011}; this lower value is more appropriate for low-metallicity galaxies as observed in our sample. The emission lines are dust corrected using a \cite{cardelli1989} attenuation prescription with $R_V = 3.1$, which is generally applicable at this redshift \citep{reddy2018}, however, as shown by \cite{sanders2024dust} the dust law may vary for individual galaxies. A more detailed description of the dust correction procedure can be found in Section \ref{sec:auroral_abund}. In case the H$\alpha$ flux is not covered by our observations, we instead use the H$\beta$ emission line and use the intrinsic Balmer flux ratio as constrained using our modelling in Section \ref{sec:auroral_abund} to convert to H$\alpha$ flux. The flux measurements are converted to luminosities using the measured redshift of each spectrum. The star formation rates of our EXCELS sample are listed in Table \ref{tab:excels_derived_data}.

\begin{table*}
\caption{The free parameters used in the measurement of auroral metallicities.}
\label{tab:auroral_params}
\centering
\begin{tabular}{llll}
\hline
Parameter & Symbol / Unit & Range & Prior   \\ \hline \hline
High ionisation zone electron temperature  & $T_e([\textsc{Oiii}])$ / K   & $(1\times10^3, 3.3\times10^4)$ & Uniform      \\
Electron density                           & $n_e$ / $\textrm{cm}^{-3}$   & $(10, 1000)$                   & Uniform      \\
Abundance of singly ionised oxygen         & O$^{+}$/H$^{+}$              & $(10^{-9}, 10^{-2})$           & Logarithmic  \\
Abundance of doubly ionised oxygen         & O$^{++}$/H$^{+}$             & $(10^{-9}, 10^{-2})$           & Logarithmic  \\
Abundance of singly ionised nitrogen       & N$^{+}$/H$^{+}$              & $(10^{-9}, 10^{-2})$           & Logarithmic  \\
$V-$band attenuation                       & $A_V$ / mag                  & $(0, 4)$                       & Uniform      \\ \hline
\end{tabular}
\end{table*}

\begin{figure}
	\begin{center}
	\includegraphics[width=0.495\textwidth]{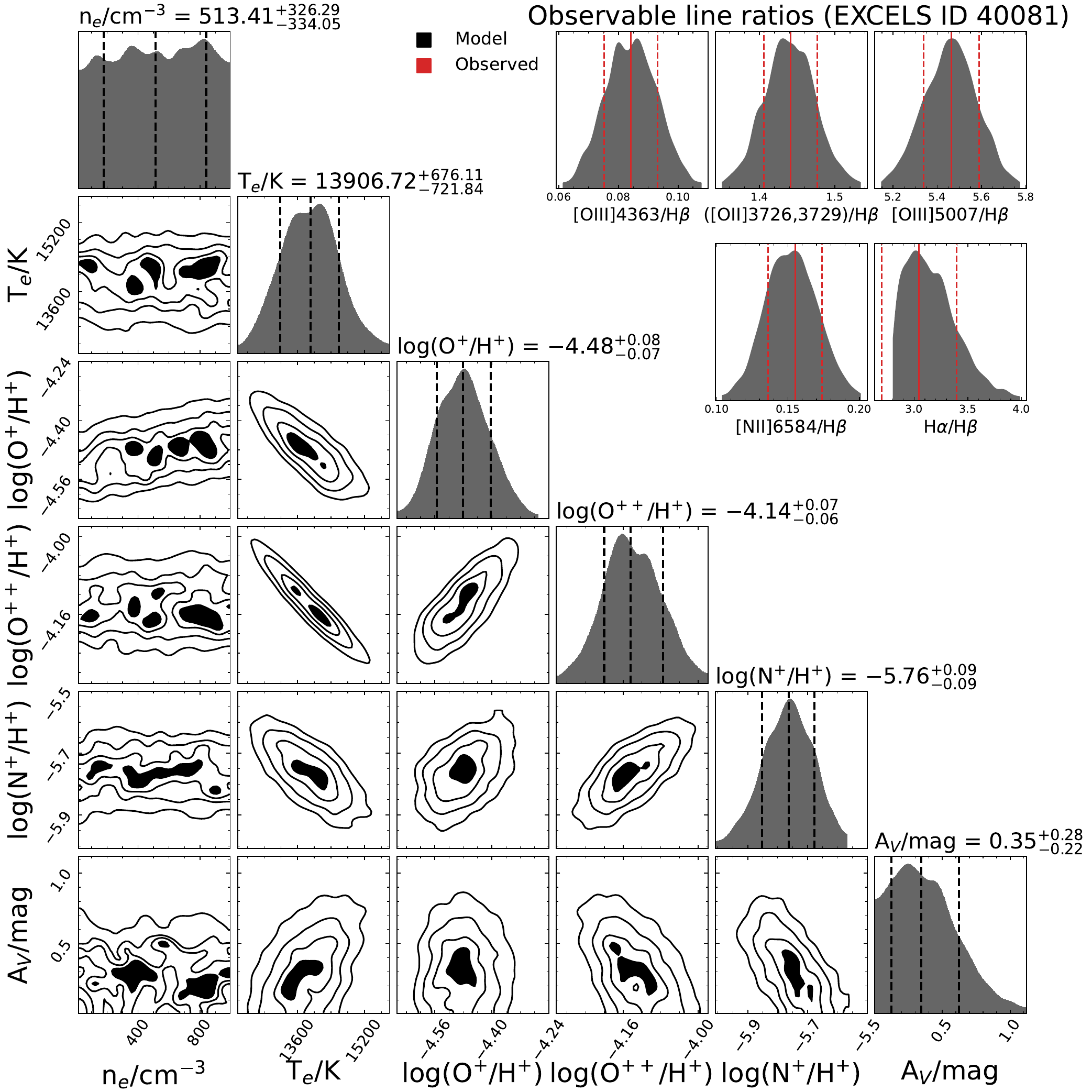}
	\caption{The posterior distribution of our abundance measurements for EXCELS target 40081. The dimensions shown are electron density ($n_e$), electron temperature ($T_e$), singly and doubly ionised oxygen (O$^{+}$/H$^{+}$ and O$^{++}$/H$^{+}$), singly ionised nitrogen (N$^{+}$/H$^{+}$) and dust attenuation ($A_V$). The observed (red) and modelled line ratios are shown in the 5 panels in the top-right corner of the figure.}
	\label{fig:corner_plot}
	\end{center}
\end{figure}

\subsection{Abundance measurements} \label{sec:auroral_abund}
We measure the gas-phase metallicities of the galaxies in our EXCELS (and DESI EDR) sample using the [\textsc{Oii}]$\lambda\lambda$3726,3729, [\textsc{Oiii}]$\lambda$4363, H$\beta$, [\textsc{Oiii}]$\lambda$5007 and H$\alpha$ or H$\gamma$ emission lines. The H$\beta$ and H$\alpha$ or H$\gamma$ emission lines are required to constrain the dust attenuation on the emission lines; we primarily use H$\alpha$ if observed and use H$\gamma$ for the remainder of EXCELS targets. Additionally, to constrain the nitrogen abundance we also use the [\textsc{Nii}]$\lambda$6584 emission line if observed. The detection of the faint [\textsc{Oiii}]$\lambda$4363 auroral emission line is crucial for this analysis as the [\textsc{Oiii}]$\lambda$4363/[\textsc{Oiii}]$\lambda$5007 ratio is sensitive to the electron temperature: $T_e([\textsc{Oiii}])$. The information on the electron temperature allows us to break the degeneracy with ionic abundances. Using the electron temperature constraint, we measure O$^{+}$/H$^{+}$, O$^{++}$/H$^{+}$ and N$^{+}$/H$^{+}$ directly from the strength of the observed emission lines.

We model the emission line fluxes using \textsc{PyNeb} \citep[][version 1.1.19]{pyneb2015} assuming a two-zone ionisation model with a high ionisation zone traced by O$^{++}$ and H$^{+}$ and a low ionisation zone traced by O$^{+}$ and N$^{+}$ \citep[][show that even for very higly ionised H\textsc{ii} regions the O$^{3+}$/H$^+$ contribution to the total oxygen abundance is negligible]{berg2021}. We use atomic and collisional data included with the \textsc{PyNeb} package (atomic: all from \cite{froesefisher2004} ; collisional: O$^{+}$ from \cite{kisielius2009}, O$^{++}$ from \cite{aggarwal1999}, N$^{+}$ from \cite{tayall2011}). We assume the temperature relation ($T_2 - T_3$) between the electron temperatures of the high and low ionisation zone as defined
\begin{equation}
    T_e([\textsc{Oii}]) = 0.7 \times T_e([\textsc{Oiii}]) + 1500,
\end{equation}
where the $T_e([\textsc{Oii}])$ and $T_e([\textsc{Oiii}])$ are given in units of Kelvin. This relation was derived by  \cite{campbell1986} and \cite{garnett1992}, however, here shown with a correction which reduces the $T_e([\textsc{Oii}])$ by 1500 K compared to the original relation\footnote{We also include a version of our measurements using the $T_e([\textsc{Oii}]) = 0.7 \times T_e([\textsc{Oiii}]) + 3000$ relation available for download. Please refer to the data availability section.}. This correction is based on the $T_2 - T_3$ relation measurements of galaxies by \cite{andrews2013} who found that $T_e([\textsc{Oii}])$ was overestimated by $\sim 1000-2000$ K; similar results were found by  \cite{pilyugin2010} and \cite{curti2017}. The nature of the discrepancy seems to be between measurements of individual \textsc{Hii}-regions and galaxy integrated measurements, where the line flux originates from a mixture of \textsc{Hii}-regions with various properties as well as diffuse ionised gas \citep[see discussion in e.g.,][]{andrews2013, curti2017, sanders2017}. Many other $T_2 - T_3$ relations have been derived, however, the relation between $T_e([\textsc{Oii}])$ and $T_e([\textsc{Oiii}])$ is poorly defined and therefore a source of systematic uncertainty in this work and others using similar techniques \citep[see e.g.,][for in depth discussions and proposed improvements]{pagel1992, izotov2006, pilyugin2009, nicholls2014, yates2020, langeroodi2024}. 

We obtained the emissivities of the various emission lines using the \texttt{getEmissivity} method of the \texttt{pn.Atom} class and \texttt{pn.RecAtom} class in the case of the Hydrogen emission line. Total line fluxes were derived by multiplying the emissivities by the ionic abundances. Lastly, to accurately model the observed emission lines we apply dust attenuation to our modelled emission lines. Here we also assume a \cite{cardelli1989} attenuation prescription with $R_V = 3.1$. We assume case-B recombination, where the intrinsic Balmer line ratios are calculated consistently according to the model electron temperature and density. To derive total metallicities we assume that the total oxygen abundance is given by
\begin{equation}
    \rm \frac{O}{H} = \frac{O^{+}}{H^{+}} + \frac{O^{++}}{H^{+}},
\end{equation}
where the contribution from higher ionisation states is negligible as shown by \cite{berg2021}. We further assume that the N/O abundance ratio is given by
\begin{equation}
    \rm \frac{N}{O} = \frac{N^{+}/H^{+}}{O^{+}/H^{+}}
\end{equation}
which provides an accurate estimation of the total N/O abundance ratio due to the similar ionisation potentials of [\textsc{Oii}] and [\textsc{Nii}] \citep[e.g.,][]{nava2006, amayo2021}.

\begin{figure*}
    \begin{center}
    \begin{subfigure}[b]{0.495\textwidth}
        \includegraphics[width=\textwidth]{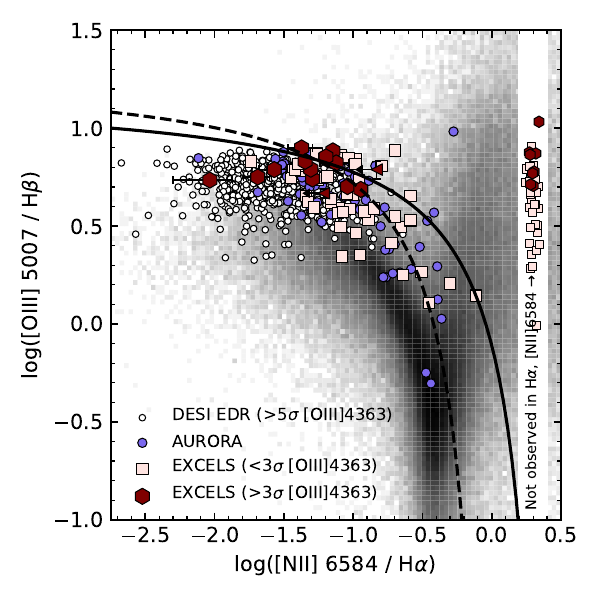}
    \end{subfigure}
    \hfill
    \begin{subfigure}[b]{0.495\textwidth}
        \includegraphics[width=\textwidth]{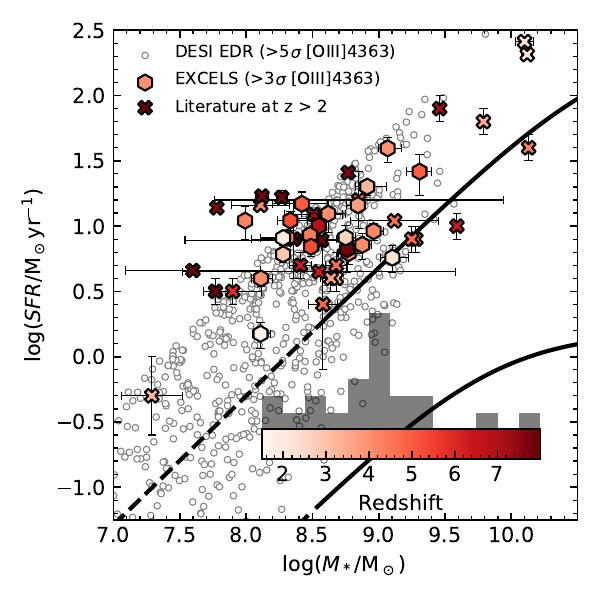}
    \end{subfigure}
    \caption{\textit{Left}: The EXCELS and DESI EDR galaxies in the [\textsc{Nii}]-BPT diagram \protect \citep[][]{baldwin1981}. We show the EXCELS (maroon-hexagons for [\textsc{Oiii}]$\lambda$4363 detections, cream squares for non-detections of [\textsc{Oiii}]$\lambda$4363), AURORA \citep[purple circles, no selection on [\textsc{Oiii}$\rbracket \lambda$4363;][]{shapley2024} and our sample of [\textsc{Oiii}]$\lambda$4363 detected DESI EDR galaxies (white circles). EXCELS galaxies which were not observed in H$\alpha$ or [\textsc{Nii}]6584 are displayed within the white band on the right side of the BPT-diagram. The full DESI EDR is shown by the grey shaded background. The solid line shows the maximum starburst line by \protect\cite{kewley2001}. The dashed line shows the \protect\cite{kauffmann2003} SF/AGN classification boundary. \textit{Right}: The EXCELS galaxies on the stellar mass versus star formation rate diagram. Our [\textsc{Oiii}]$\lambda$4363 detected EXCELS galaxies are shown using hexagons. Literature [\textsc{Oiii}]$\lambda$4363 detected galaxies are shown using crosses. All the data are coloured by their redshift as indicated by the colour bar. The histogram above the colour bar shows the redshift distribution of our EXCELS sample. The black lines show the average star formation main sequence of galaxies at $z=0$ (bottom) and $z=4$ (top) as measured by \protect \cite{popesso2023}. We note that the DESI star formation rates are derived using SED fitting (see Section \ref{sec:desi_data}).}
    \label{fig:general_properties}
    \end{center}
\end{figure*}

We use a forward modelling approach to simultaneously constrain all the modelled parameters. We use the \textsc{dynesty} \citep{dynesty2020, dynesty2022} dynamic nested sampling code to explore the parameter space. Our model has 6 free parameters ($\bm \theta$): $T_e([\textsc{Oiii}])$, $n_e$, log(O$^{+}$/H$^{+}$), log(O$^{++}$/H$^{+}$), log(N$^{+}$/H$^{+}$) and $A_{v}$. We assume a (log-)uniform prior (${\bm \pi}({\bm \theta})$) on all parameters (see Table \ref{tab:auroral_params}). Almost all parameters are well constrained by the observations. The exception to this is the measurement of the electron density, which is unconstrained within the prior bounds between 10 cm$^{-3}$ and 1000 cm$^{-3}$. However, the other parameters are not strongly correlated with electron density\footnote{Our $T_e$ measurements and derived abundances are insensitive to the electron density where $n_e \lesssim 10^4 ~\rm cm^{-3}$; higher densities are not expected for the sample probed here \citep[e.g.][]{isobe2023,reddy2023sfr_ne}.}. We assume a Gaussian likelihood function. The posterior function is given by
\begin{equation}
    \begin{split}
    \textrm{ln}~ & p({\bm R}_{\rm obs}, {\bm \sigma}_{\rm obs}|{\bm \theta}) \propto \textrm{ln}~ {\bm \pi}({\bm \theta}) \\ 
    & -\frac{1}{2} \sum \left[ \frac{\{{\bm R}_{\rm obs} - {\bm R}_{\rm model}({\bm \theta})\}^{2}}{{\bm \sigma}_{\rm obs}^{2}} \right],
    \end{split}
\end{equation}
where ${\bm R}_{\rm obs}$ and ${\bm \sigma}_{\rm obs}$ are the observed flux ratio and associated uncertainty of the [\textsc{Oii}]$\lambda\lambda$3726,3729, [\textsc{Oiii}]$\lambda$4363, [\textsc{Oiii}]$\lambda$5007, [\textsc{Nii}]$\lambda$6584 and H$\alpha$ (or H$\gamma$) emission lines relative to H$\beta$. ${\bm R}_{\rm model}({\bm \theta})$ are the modelled line flux ratios of these same emission lines. An example posterior distribution of an EXCELS target is shown in Figure \ref{fig:corner_plot}. The measured constraints from this modelling are shown for the EXCELS and DESI EDR samples in Tables \ref{tab:excels_derived_data} and \ref{tab:desi_derived_data}, respectively.

The forward modelling approach used here differs from the method traditionally used to infer abundances using \textsc{PyNeb} \citep{pyneb2015} or similar methods in \textsc{iraf/stsdas} \citep{shaw1995}. Typically, a multi-step approach is used. First, the emission lines are dust corrected. Following this, electron temperatures (and densities) are constrained using the \texttt{getTemDen} (or \texttt{getCrossTemDen}) routine. Then the ionic abundances of different elements are constrained using \texttt{getIonAbundance}. The uncertainties on the derived quantities are obtained using Monte Carlo simulations where the inputted line ratios are varied within the measurement uncertainties. Our approach yields consistent results with the traditional approach \citep[e.g.][]{stanton2024excels, arellano-cordova2024excels}\footnote{All the measurements are statistically consistent. There may be some minor differences due to different assumptions on the dust attenuation prescription, temperature relations and used atomic and collisional data.}. 

There are several advantages to the forward modelling approach. A good example of this is the calculation of the intrinsic case-B recombination fluxes of the Balmer lines which are used to dust-correct the measurements of all the emission lines. We calculate the intrinsic Balmer ratios for each model evaluation. This is important as the intrinsic Balmer ratios between H$\alpha$/H$\beta$ and H$\gamma$/H$\beta$ are dependent on the electron temperature. For example, metal poor dwarf with $T_{e}(\textsc{Oiii}) = 2.5 \times 10^4~\rm K$ has a case-B H$\alpha$/H$\beta$ flux ratio of $\sim2.72$ as compared to the canonical value of 2.86 (at $T_{e}(\textsc{Oiii}) \sim 1 \times 10^4~\rm K$) that is typically used for dust corrections. Our method also explicitly permits the marginalisation over unconstrained (nuisance) parameters in our measurements, e.g., electron density in our current analysis.

\renewcommand{\arraystretch}{1.25}
\begin{table*}
\centering
\caption{Our derived data products for the EXCELS sample. These include the EXCELS target IDs, redshifts, stellar masses and star formation rates derived using \textsc{Bagpipes}, H$\alpha$ star formation rates, and our constraints on the dust attenuation, electron temperature and elemental abundances derived using our forward modelled \textsc{PyNeb} implementation. The full table is available in machine readable format (see data availability section).}
\begin{tabular}{ccccccccccc}
\hline

Target ID & Redshift & log($M_{\star}$) & \textit{SFR$_{\rm SED}$} & \textit{SFR$_{\rm H\alpha|H\beta}$} & $A_V$ & $T_e$ & $\rm 12 + log(O/H)$ & $\rm log(O^{+}/H^{+})$ & $\rm log(O^{++}/H^{+})$ & $\rm log(N/O)$ \\
 & & log(M$_{\odot}$) & M$_{\odot}$ yr$^{-1}$ & M$_{\odot}$ yr$^{-1}$ & mag & $\times 10^4$ K & & & & \\
\hline \hline
40081 & 3.952 & ${9.06}_{-0.07}^{+0.10}$ & ${33.6}_{-8.40}^{+11.5}$ & ${39.4}_{-8.53}^{+8.53}$ & ${0.34}_{-0.22}^{+0.27}$ & ${1.39}_{-0.07}^{+0.06}$ & ${8.02}_{-0.05}^{+0.06}$ & ${-4.48}_{-0.07}^{+0.07}$ & ${-4.14}_{-0.05}^{+0.06}$ & ${-1.28}_{-0.09}^{+0.08}$ \\ 
45177 & 2.899 & ${8.27}_{-0.05}^{+0.05}$ & ${16.5}_{-3.12}^{+3.39}$ & ${6.08}_{-0.80}^{+0.80}$ & ${0.39}_{-0.09}^{+0.08}$ & ${1.40}_{-0.13}^{+0.12}$ & ${8.10}_{-0.10}^{+0.13}$ & ${-4.50}_{-0.13}^{+0.16}$ & ${-4.01}_{-0.09}^{+0.12}$ & ${-1.10}_{-0.08}^{+0.08}$ \\ 
45393 & 4.234 & ${8.91}_{-0.08}^{+0.14}$ & ${35.1}_{-13.1}^{+19.6}$ & ${11.0}_{-3.12}^{+3.12}$ & ${0.33}_{-0.21}^{+0.28}$ & ${1.37}_{-0.13}^{+0.12}$ & ${8.03}_{-0.11}^{+0.13}$ & ${-4.38}_{-0.13}^{+0.16}$ & ${-4.16}_{-0.10}^{+0.13}$ & --- \\ 
47557 & 3.233 & ${8.54}_{-0.03}^{+0.03}$ & ${23.5}_{-4.20}^{+4.81}$ & ${20.0}_{-2.77}^{+2.77}$ & ${1.04}_{-0.10}^{+0.10}$ & ${1.50}_{-0.18}^{+0.19}$ & ${7.95}_{-0.14}^{+0.16}$ & ${-4.57}_{-0.18}^{+0.19}$ & ${-4.19}_{-0.12}^{+0.15}$ & ${-1.26}_{-0.10}^{+0.10}$ \\ 
48659 & 6.795 & ${8.61}_{-0.13}^{+0.13}$ & ${6.24}_{-2.23}^{+3.78}$ & ${10.0}_{-3.10}^{+3.10}$ & ${0.84}_{-0.37}^{+0.36}$ & ${2.52}_{-0.44}^{+0.25}$ & ${7.36}_{-0.08}^{+0.16}$ & ${-6.00}_{-1.53}^{+0.37}$ & ${-4.65}_{-0.09}^{+0.15}$ & --- \\ 
52422 & 4.022 & ${8.11}_{-0.06}^{+0.08}$ & ${5.37}_{-1.20}^{+1.77}$ & ${12.4}_{-3.53}^{+3.53}$ & ${0.70}_{-0.34}^{+0.32}$ & ${2.13}_{-0.20}^{+0.21}$ & ${7.47}_{-0.08}^{+0.09}$ & ${-5.44}_{-0.12}^{+0.12}$ & ${-4.57}_{-0.08}^{+0.09}$ & ${-1.67}_{-0.86}^{+0.33}$ \\ 
56875 & 3.997 & ${8.84}_{-0.08}^{+0.05}$ & ${67.3}_{-11.3}^{+9.77}$ & ${3.97}_{-0.51}^{+0.51}$ & ${0.06}_{-0.04}^{+0.09}$ & ${1.93}_{-0.31}^{+0.31}$ & ${7.57}_{-0.14}^{+0.19}$ & ${-5.00}_{-0.19}^{+0.23}$ & ${-4.55}_{-0.13}^{+0.18}$ & --- \\ 
57498 & 3.693 & ${8.48}_{-0.03}^{+0.04}$ & ${25.4}_{-4.00}^{+4.87}$ & ${14.4}_{-1.75}^{+1.75}$ & ${0.44}_{-0.05}^{+0.05}$ & ${1.47}_{-0.06}^{+0.05}$ & ${7.99}_{-0.04}^{+0.05}$ & ${-4.61}_{-0.06}^{+0.07}$ & ${-4.12}_{-0.04}^{+0.05}$ & ${-1.24}_{-0.04}^{+0.04}$ \\ 
59009 & 4.132 & ${8.31}_{-0.06}^{+0.10}$ & ${10.0}_{-2.69}^{+2.92}$ & ${9.14}_{-1.39}^{+1.39}$ & ${0.09}_{-0.07}^{+0.12}$ & ${1.29}_{-0.06}^{+0.05}$ & ${8.17}_{-0.05}^{+0.06}$ & ${-4.53}_{-0.07}^{+0.08}$ & ${-3.91}_{-0.05}^{+0.06}$ & --- \\ 
59720 & 4.365 & ${9.30}_{-0.09}^{+0.09}$ & ${16.4}_{-3.29}^{+5.63}$ & ${8.54}_{-1.28}^{+1.28}$ & ${0.12}_{-0.09}^{+0.17}$ & ${1.26}_{-0.13}^{+0.14}$ & ${8.12}_{-0.14}^{+0.17}$ & ${-4.30}_{-0.16}^{+0.20}$ & ${-4.07}_{-0.12}^{+0.15}$ & ${-1.17}_{-0.11}^{+0.09}$ \\ 
63962 & 4.356 & ${8.48}_{-0.04}^{+0.07}$ & ${21.0}_{-4.68}^{+4.29}$ & ${8.15}_{-0.95}^{+0.95}$ & ${0.03}_{-0.02}^{+0.05}$ & ${1.79}_{-0.21}^{+0.20}$ & ${7.66}_{-0.11}^{+0.13}$ & ${-5.21}_{-0.14}^{+0.17}$ & ${-4.39}_{-0.10}^{+0.12}$ & ${-1.40}_{-0.15}^{+0.12}$ \\ 
69991 & 4.935 & ${8.28}_{-0.16}^{+0.18}$ & ${1.98}_{-0.69}^{+1.82}$ & ${26.1}_{-9.08}^{+9.08}$ & ${2.17}_{-0.44}^{+0.39}$ & ${1.58}_{-0.27}^{+0.30}$ & ${7.96}_{-0.19}^{+0.23}$ & ${-4.46}_{-0.22}^{+0.28}$ & ${-4.23}_{-0.18}^{+0.21}$ & ${-1.28}_{-0.23}^{+0.17}$ \\ 
70864 & 5.252 & ${8.33}_{-0.09}^{+0.12}$ & ${8.21}_{-2.23}^{+4.19}$ & ${6.95}_{-1.14}^{+1.14}$ & ${0.16}_{-0.12}^{+0.20}$ & ${1.26}_{-0.12}^{+0.12}$ & ${8.20}_{-0.12}^{+0.14}$ & ${-4.60}_{-0.14}^{+0.18}$ & ${-3.87}_{-0.11}^{+0.14}$ & ${-1.21}_{-0.15}^{+0.12}$ \\ 
73535 & 2.208 & ${8.77}_{-0.08}^{+0.10}$ & ${11.4}_{-3.71}^{+6.02}$ & ${5.72}_{-1.34}^{+1.34}$ & ${0.24}_{-0.16}^{+0.22}$ & ${1.08}_{-0.13}^{+0.11}$ & ${8.40}_{-0.14}^{+0.21}$ & ${-4.02}_{-0.17}^{+0.24}$ & ${-3.79}_{-0.13}^{+0.19}$ & --- \\ 
93897 & 4.079 & ${7.99}_{-0.02}^{+0.02}$ & ${10.1}_{-0.47}^{+0.61}$ & ${7.18}_{-1.59}^{+1.59}$ & ${0.14}_{-0.10}^{+0.23}$ & ${1.44}_{-0.11}^{+0.11}$ & ${8.21}_{-0.11}^{+0.11}$ & ${-4.44}_{-0.12}^{+0.13}$ & ${-3.88}_{-0.11}^{+0.10}$ & --- \\ 
94335 & 1.810 & ${8.96}_{-0.03}^{+0.07}$ & ${39.6}_{-14.7}^{+16.4}$ & ${8.10}_{-0.96}^{+0.96}$ & ${0.57}_{-0.04}^{+0.04}$ & ${1.11}_{-0.07}^{+0.07}$ & ${8.31}_{-0.09}^{+0.11}$ & ${-4.05}_{-0.10}^{+0.13}$ & ${-3.93}_{-0.08}^{+0.10}$ & ${-1.12}_{-0.05}^{+0.05}$ \\ 
95839 & 4.955 & ${9.10}_{-0.09}^{+0.12}$ & ${20.7}_{-5.27}^{+6.77}$ & ${14.8}_{-3.43}^{+3.43}$ & ${0.26}_{-0.16}^{+0.21}$ & ${1.42}_{-0.06}^{+0.06}$ & ${8.05}_{-0.05}^{+0.05}$ & ${-4.64}_{-0.06}^{+0.07}$ & ${-4.04}_{-0.05}^{+0.05}$ & --- \\ 
104937 & 1.651 & ${8.87}_{-0.06}^{+0.06}$ & ${1.88}_{-0.31}^{+0.33}$ & ${1.49}_{-0.33}^{+0.33}$ & ${0.17}_{-0.13}^{+0.23}$ & ${1.24}_{-0.16}^{+0.16}$ & ${8.24}_{-0.16}^{+0.21}$ & ${-4.12}_{-0.19}^{+0.24}$ & ${-3.99}_{-0.15}^{+0.18}$ & --- \\ 
119504 & 7.916 & ${8.42}_{-0.22}^{+0.19}$ & ${7.92}_{-2.50}^{+4.71}$ & ${6.63}_{-2.23}^{+2.23}$ & ${0.28}_{-0.21}^{+0.42}$ & ${2.16}_{-0.33}^{+0.37}$ & ${7.57}_{-0.14}^{+0.16}$ & ${-5.50}_{-0.19}^{+0.21}$ & ${-4.46}_{-0.14}^{+0.16}$ & --- \\ 
121806 & 5.226 & ${8.10}_{-0.06}^{+0.07}$ & ${11.7}_{-2.01}^{+2.73}$ & ${11.0}_{-1.80}^{+1.80}$ & ${0.17}_{-0.12}^{+0.19}$ & ${1.51}_{-0.10}^{+0.11}$ & ${7.95}_{-0.08}^{+0.08}$ & ${-4.86}_{-0.10}^{+0.11}$ & ${-4.12}_{-0.08}^{+0.08}$ & ${-0.93}_{-0.10}^{+0.08}$ \\ 
123597 & 3.797 & ${8.76}_{-0.05}^{+0.07}$ & ${29.9}_{-11.4}^{+13.2}$ & ${6.21}_{-1.11}^{+1.11}$ & ${0.20}_{-0.14}^{+0.23}$ & ${1.15}_{-0.14}^{+0.13}$ & ${8.28}_{-0.15}^{+0.20}$ & ${-4.23}_{-0.18}^{+0.23}$ & ${-3.86}_{-0.14}^{+0.19}$ & ${-1.67}_{-0.17}^{+0.14}$ \\ 
123837 & 2.615 & ${8.75}_{-0.06}^{+0.07}$ & ${8.85}_{-1.84}^{+2.75}$ & ${8.17}_{-1.91}^{+1.91}$ & ${0.18}_{-0.13}^{+0.26}$ & ${1.76}_{-0.18}^{+0.19}$ & ${7.69}_{-0.10}^{+0.12}$ & ${-4.91}_{-0.13}^{+0.15}$ & ${-4.42}_{-0.09}^{+0.11}$ & --- \\ \hline
\end{tabular}
\label{tab:excels_derived_data}
\end{table*}
\renewcommand{\arraystretch}{1.}

\renewcommand{\arraystretch}{1.25}
\begin{table*}
\centering
\caption{Our derived data products for the DESI EDR sample. We provide the DESI identifiers required for the identification of unique spectra (target ID, survey, program and healpix) and our constraints on the dust attenuation, electron temperature and elemental abundances derived using our forward modelled \textsc{PyNeb} implementation. Only the first ten rows of this table are included here for illustration. The full table is available in machine readable format (see data availability section).}
\begin{tabular}{cccccccccc}
\hline
Target ID & Survey & Program & Healpix & $A_V$ & $T_e$ & $\rm 12 + log(O/H)$ & $\rm log(O^{+}/H^{+})$ & $\rm log(O^{++}/H^{+})$ & $\rm log(N/O)$ \\
 & & & & mag & $\times 10^4$ K & & &\\
\hline \hline
49749096923136 & sv1 & dark & 26280 & ${0.09}_{-0.05}^{+0.05}$ & ${1.71}_{-0.17}^{+0.18}$ & ${7.58}_{-0.11}^{+0.13}$ & ${-4.74}_{-0.12}^{+0.15}$ & ${-4.69}_{-0.09}^{+0.11}$ & ${-1.39}_{-0.06}^{+0.06}$ \\ 
103633278468120 & sv1 & other & 11318 & ${0.33}_{-0.07}^{+0.06}$ & ${1.77}_{-0.18}^{+0.19}$ & ${7.71}_{-0.11}^{+0.12}$ & ${-4.72}_{-0.13}^{+0.14}$ & ${-4.48}_{-0.10}^{+0.11}$ & ${-1.77}_{-0.13}^{+0.10}$ \\ 
234514861260805 & sv1 & other & 27250 & ${0.05}_{-0.03}^{+0.04}$ & ${1.78}_{-0.11}^{+0.12}$ & ${7.61}_{-0.07}^{+0.07}$ & ${-5.15}_{-0.09}^{+0.09}$ & ${-4.46}_{-0.06}^{+0.06}$ & ${-1.39}_{-0.10}^{+0.08}$ \\ 
234520867504135 & sv1 & other & 27247 & ${0.05}_{-0.03}^{+0.04}$ & ${1.87}_{-0.15}^{+0.16}$ & ${7.59}_{-0.09}^{+0.09}$ & ${-4.81}_{-0.10}^{+0.11}$ & ${-4.62}_{-0.08}^{+0.08}$ & ${-1.41}_{-0.06}^{+0.05}$ \\ 
39627152907306205 & sv1 & dark & 38777 & ${0.20}_{-0.06}^{+0.06}$ & ${1.60}_{-0.12}^{+0.12}$ & ${7.78}_{-0.08}^{+0.09}$ & ${-5.12}_{-0.10}^{+0.12}$ & ${-4.27}_{-0.07}^{+0.08}$ & ${-1.12}_{-0.08}^{+0.08}$ \\ 
39627158280212276 & sv1 & dark & 38778 & ${0.12}_{-0.02}^{+0.02}$ & ${1.25}_{-0.05}^{+0.05}$ & ${8.20}_{-0.06}^{+0.07}$ & ${-3.98}_{-0.07}^{+0.07}$ & ${-4.26}_{-0.05}^{+0.05}$ & ${-1.48}_{-0.03}^{+0.03}$ \\ 
39627163716027958 & sv1 & dark & 38780 & ${0.02}_{-0.01}^{+0.03}$ & ${1.83}_{-0.15}^{+0.14}$ & ${7.56}_{-0.08}^{+0.09}$ & ${-4.93}_{-0.10}^{+0.11}$ & ${-4.59}_{-0.07}^{+0.08}$ & ${-1.70}_{-0.13}^{+0.11}$ \\ 
39627179973149339 & sv1 & dark & 38783 & ${0.79}_{-0.03}^{+0.03}$ & ${1.09}_{-0.06}^{+0.06}$ & ${8.35}_{-0.09}^{+0.10}$ & ${-3.92}_{-0.10}^{+0.11}$ & ${-3.97}_{-0.08}^{+0.09}$ & ${-1.32}_{-0.04}^{+0.04}$ \\ 
39627185329276918 & sv1 & dark & 38863 & ${0.19}_{-0.03}^{+0.03}$ & ${1.53}_{-0.04}^{+0.04}$ & ${7.91}_{-0.03}^{+0.03}$ & ${-4.93}_{-0.04}^{+0.04}$ & ${-4.15}_{-0.02}^{+0.03}$ & ${-1.68}_{-0.15}^{+0.11}$ \\ 
39627190802845790 & sv1 & dark & 38872 & ${0.41}_{-0.02}^{+0.02}$ & ${1.21}_{-0.04}^{+0.04}$ & ${8.25}_{-0.04}^{+0.05}$ & ${-4.12}_{-0.06}^{+0.06}$ & ${-3.98}_{-0.04}^{+0.04}$ & ${-1.54}_{-0.03}^{+0.03}$ \\ \multicolumn{10}{c}{...} \\
\hline
\end{tabular}
\label{tab:desi_derived_data}
\end{table*}
\renewcommand{\arraystretch}{1.}

\begin{figure*}
	\begin{center}
	\includegraphics[width=\textwidth, trim=0 10.3cm 0 0, clip]{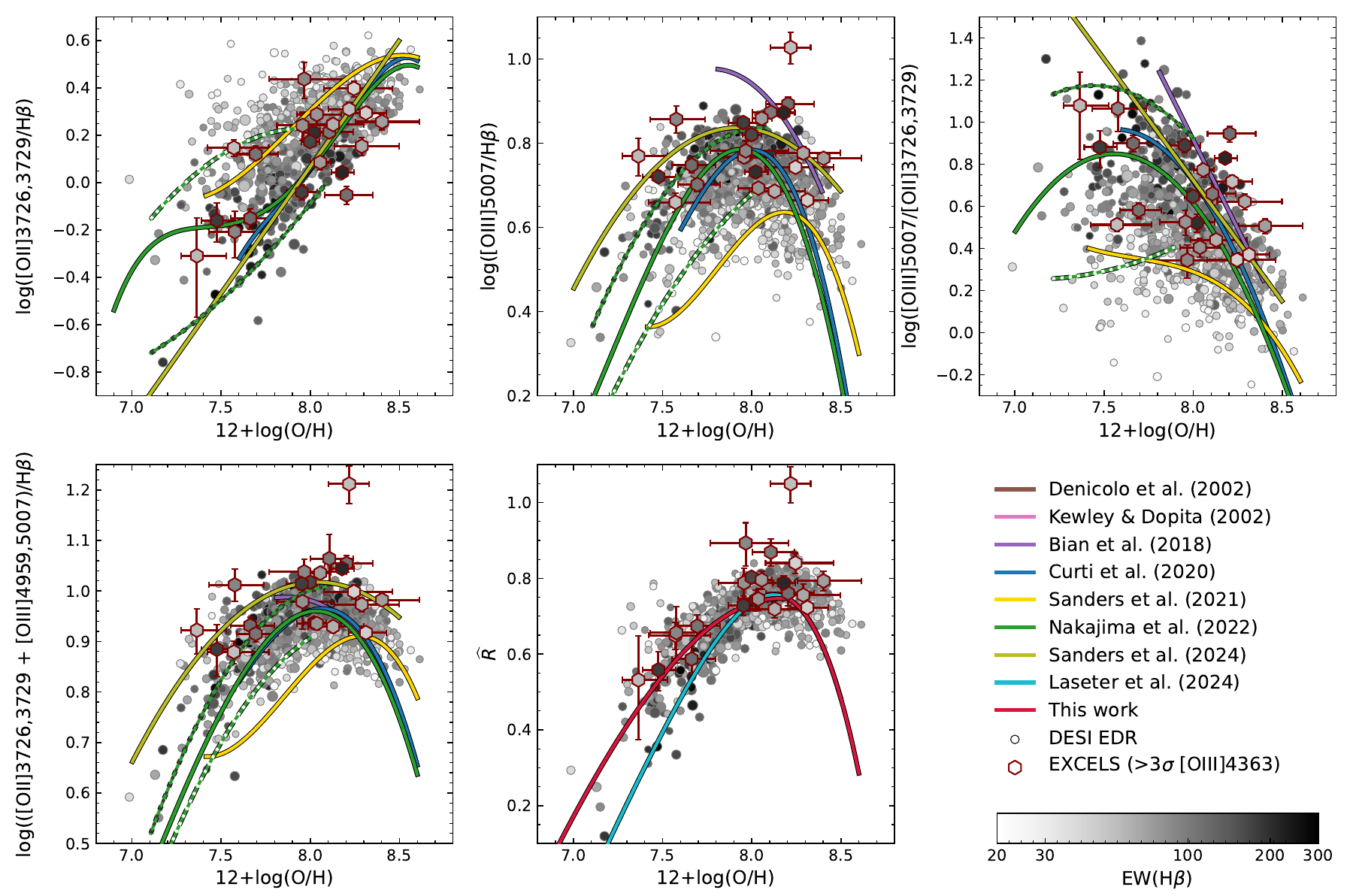}
	\caption{The relation between $T_e$-metallicity and strong-line ratios of the \textit{JWST} EXCELS (\textit{hexagons, maroon edges}) and the local sample of DESI EDR galaxies (\textit{circles, grey edges}). The \textit{R2}, \textit{R3}, \textit{O32}, \textit{R23}, $\widehat{R}$, (\textit{left-to-right} and \textit{top-to-bottom}) strong-line ratios are shown in the individual panels. The data points are coloured according to their \textrm{H}$\beta$ equivalent width as described by the colourbar. Together with the measured data points we show several literature strong-line metallicity calibrations by \protect\cite{bian2018, curti2020, sanders2021, nakajima2022, sanders2024} and \protect\cite{laseter2024}. The dashed green-white and green-black lines show the \protect\cite{nakajima2022} low-EW(H$\beta$) and high-EW(H$\beta$) metallicity calibrations, respectively. The legend describes the calibrations displayed in all of the emission line diagrams in this work. The legend therefore includes some calibrations which are used in one of the other figures. The legend is applicable to Figures \ref{fig:Rhat_calibration}, \ref{fig:auroral_lineratios_hb_ew_nitrogen}, \ref{fig:auroral_lineratios_hb_ew_neon_sulfur}, \ref{fig:RNehat_gamma_calibration} and \ref{fig:rhat_chakraborty}.}
	\label{fig:auroral_lineratios_hb_ew_oxygen}
	\end{center}
\end{figure*}

\section{General properties of the sample} \label{sec:properties}
As a result of survey design and strict selection criteria (e.g., an [\textsc{Oiii}]$\lambda$4363 emission line detection) the sample of galaxies studied here cannot be considered a representative sample of the galaxy population at their respective redshift. This bias is present in the galaxy sample we compiled from the EXCELS survey as well as the DESI EDR local comparison sample. We are particularly biased to low-metallicity, highly star-forming galaxies. To highlight the galaxy populations captured by our selections, we show the location of these galaxies on the [\textsc{Nii}]-BPT emission line diagram \citep{baldwin1981} on the left panel of Figure \ref{fig:general_properties}. We show the full galaxy population traced by the DESI EDR in the grey shaded background. We show the [\textsc{Oiii}]$\lambda$4363 detected EXCELS and DESI EDR galaxies in maroon hexagons and white circles, respectively. We also show the EXCELS galaxies which are not included in our sample (cream squares) and galaxies observed in the AURORA survey \citep{shapley2024}. In the right panel of Figure \ref{fig:general_properties}, we show the location of our EXCELS galaxies on the star-formation main sequence \citep[e.g.,][]{brinchmann2004, speagle2014, saintonge2017, popesso2023, clarke2024}. This figure shows that the galaxies in our EXCELS and DESI EDR samples consistently have higher star formation rates compared to the general population at their respective redshift \citep[as shown by the average $z$-dependent main sequence of][black lines]{popesso2023}. The median offset of our EXCELS galaxies from the \cite{popesso2023} main sequence is 0.7 dex. In this Figure, we also include a selection of measurements taken from the literature \citep[see Section \ref{sec:literature_sample}, data taken from][]{sanders2023metallicity,curti2023, nakajima2023, morishita2024} which occupy a very similar parameter space to the galaxies presented here. We add the caveat that the DESI star formation rates are derived inhomogeneously from the rest of our measurements using SED fitting instead of H$\alpha$ flux.

\section{strong-line calibrations at high-redshift} \label{sec:strong_lines}
The emission line ratios of many strong emission lines in the rest-frame optical spectrum correlate with metallicity but also the ionisation parameter, ISM pressure and other element abundances \citep[for a review see][]{kewley2019review}. In fact, some of the success of strong-line calibrations is due to the systematic evolution of these secondary dependencies as a function of the metallicity of galaxies. However, this also makes strong-line calibrations susceptible to bias when the properties of the inferred sample deviate significantly from the calibration sample. Due to the significant evolution of the ISM conditions of the galaxy population as a function of redshift \citep{steidel2014, shapley2015, sanders2016, kashino2017, strom2017, strom2018, topping2020, cullen2021, kashino2022, sanders2023excitation, shapley2024, chartab2024, stanton2024kmos, arellano-cordova2024excels}, it is important that we verify that strong-line diagnostics can effectively recover the metallicities of galaxies over all epochs.

We use our sample of high-redshift and local $T_e$-metallicity measurements to assess the success of various strong-line calibrations at predicting metallicities for these samples. We will first investigate strong-line calibrations purely based on oxygen and hydrogen emission lines and following this also assess the success of calibrations including emission lines of different elements, particularly nitrogen, neon and sulfur. We investigate the effect of variations in the ionising radiation using the measured rest-frame H$\beta$ equivalent width which strongly correlates with ionisation parameter \citep[e.g.,][]{dottori1981, nakajima2022}. 

Throughout this section we do not describe in detail which particular calibrations are the most reliable to estimate metallicities using each set of line ratios. However, we do include the residuals of each of the discussed calibrations on our EXCELS and DESI EDR samples in Appendix \ref{sec:appendix_residuals_diagnostics}.

\begin{figure*}
	\begin{center}
	\includegraphics[width=\textwidth]{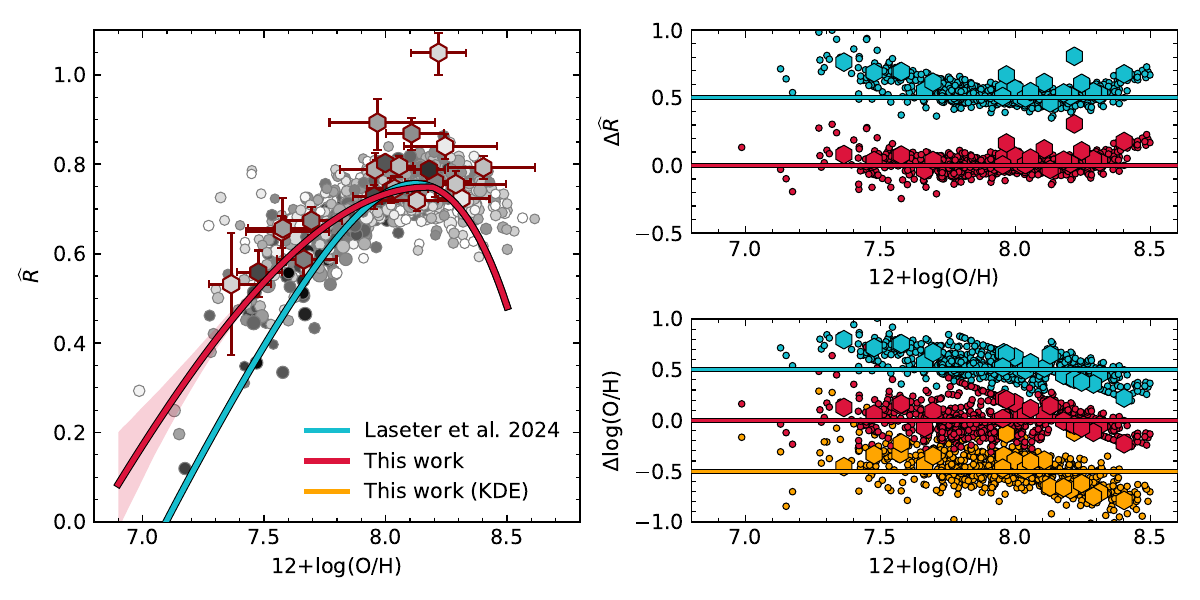}
	\caption{\textit{Left}: An enlarged version of the $\widehat{R}$ panel in Figure \ref{fig:auroral_lineratios_hb_ew_oxygen} where the data points are still coloured using the same colourbar. In addition to the $\widehat{R}$ calibration by \protect\cite{laseter2024} we also show our own calibration based on the DESI EDR data (Eq. \ref{eq:rhat_this_work}, red line) along with the 1-$\sigma$ fit uncertainties shaded in light pink. \textit{Top-right}: A comparison of the residuals in the $\Delta \widehat{R}$ direction derived using the $\widehat{R}$ calibration by \protect\cite{laseter2024} in cyan and our $\widehat{R}$ calibration in Equation \ref{eq:rhat_this_work} in red. \textit{Bottom-right}: Same as top-right, however, here showing residuals in the $\Delta$log(O/H) direction. Here we also show the residuals derived using the non-parametric KDE method in orange. See Section \ref{sec:rhat_recalibration} and Appendix \ref{sec:appendix_residuals_diagnostics} for further details on the residuals of each calibration.}
	\label{fig:Rhat_calibration}
	\end{center}
\end{figure*}

\subsection{Oxygen line based calibrations} \label{sec:oxygen_calibrators}
First, we look at strong-line calibrations based purely on emission lines from oxygen and hydrogen. In Figure \ref{fig:auroral_lineratios_hb_ew_oxygen} we compare the relation between our $T_e$-metallicities compared to the $R2$, $R3$, $O32$, $R23$ and $\widehat{R}$ strong-line ratios as defined in Section \ref{sec:intro}. Overlaid onto these figures we show strong-line diagnostics derived by \cite{bian2018, curti2020, sanders2021, nakajima2022, sanders2024} and \cite{laseter2024}. We chose these particular strong-line diagnostics as they are frequently used and based entirely on $T_e$-metallicity measurements. The data points are coloured according to their EW(H$\beta$), which is a proxy of the ionisation parameter of the radiation impacting the gas in the star-forming regions of these galaxies. 

The $R2$ and $R3$ diagnostics both use simple oxygen-to-Balmer line ratios to infer metallicities. The key difference between these calibrations is that the $R2$ diagnostic uses the [\textsc{Oii}]$\lambda\lambda$ 3727,3729 line of singly ionised oxygen and $R3$ uses the [\textsc{Oiii}]$\lambda$ 5007 line produced by doubly ionised oxygen. Both metallicity diagnostics have an inherent secondary dependence on ionisation parameter; i.e.\ higher ionisation parameter star-forming regions produce lower/higher $R2$/$R3$ ratios at fixed metallicity as these indicators trace different ionisation states. This effect is visible in Figure \ref{fig:auroral_lineratios_hb_ew_oxygen} through the clear gradients in EW(H$\beta$). 
The scatter in Figure \ref{fig:auroral_lineratios_hb_ew_oxygen} shows that the dependence on ionisation parameter should be addressed when mapping between line ratios and metallicity \citep[see also e.g.,][]{kewley2002, nakajima2022}. 
We find a similar result for the $O32$ diagnostic. The $O32$ ratio only has a weak inherent dependence on metallicity; the main dependence of this line ratio is on the strength of the ionising radiation \citep[][]{kewley2002}. Therefore, the correlation of the $O32$ diagnostic with metallicity is largely due to the evolving nature of the average strength of ionising radiation in star-forming regions as a function of the gas-phase metallicity of galaxies. Several methods have been used to address the dependency on ionisation parameter. One method is to explicitly include ionisation parameter (or a proxy such as EW(H$\beta$)) in the calibration \citep[e.g.,][]{nakajima2022} or to carefully match the ionisation parameter range of the calibration sample to the sample for which you infer metallicities; examples of this approach are local analogue samples used to construct strong-line calibrations for high-redshift galaxies \citep[][]{bian2018, sanders2021} or strong-line calibrations derived using high-redshift galaxies and applied to the same regime \citep[e.g.][]{sanders2024}.

There are two oxygen line diagnostics which show less dependence on ionisation parameter, $R23$ and $\widehat{R}$. Due to the inclusion of emission lines of both singly and doubly ionised oxygen in the numerator, the $R23$ diagnostic has a reduced (but not eliminated) sensitivity to ionisation parameter. This generally reliable diagnostic has been commonly used since its inception \citep{pagel1979}. However, there is a large plateau around $\rm 12+\log(OH) \sim8.$ where the $R23$ ratio is largely insensitive to metallicity. The $\widehat{R}$ diagnostic introduced by \cite{laseter2024} further minimises the secondary dependence on ionisation parameter using an optimal weighting between the contribution of emission from the [\textsc{Oii}]$\lambda\lambda$3727,3729 and [\textsc{Oiii}]$\lambda$5007 lines. This weighting significantly reduces the scatter for this diagnostic and shrinks the metallicity range where this diagnostic is insensitive to metallicity due to a sharper turnover than the $R23$ diagnostic, as shown in Figure \ref{fig:auroral_lineratios_hb_ew_oxygen}. Almost all local and high-redshift galaxies are co-located onto a narrow sequence in the plane spanned by metallicity and the $\widehat{R}$ line ratio. Our results further demonstrate the success of the approach used by \cite{laseter2024} to find an optimal projection of line ratios which minimises secondary dependencies beyond metallicity. We also find excellent consistency between the EXCELS high-redshift galaxies and our local DESI EDR comparison sample. However, we note that the narrow sequence traced by the galaxies in both our local and high-redshift samples deviates from the \cite{laseter2024} calibration at the lowest metallicities. At fixed $\widehat{R}$ we systematically measure lower $T_e$-metallicities in the low metallicity regime. The origin of this discrepancy between the relation we find and \cite{laseter2024} is unclear; it may be due to sample selection effects or differences in the analysis methodology. If the difference is due to sample selection this may imply that the true scatter in $\widehat{R}$ is larger than implied by current calibrations. Further work with larger and more complete samples is required to understand the discrepancies. Whilst the $\widehat{R}$ diagnostic shows a clear metallicity dependence which is independent of ionisation parameter, there is still a region at $\widehat{R}\gtrsim0.7$ where this diagnostic is largely insensitive to metallicities between $7.8 \lesssim 12+\rm log(O/H) \lesssim 8.3$; due to this there is a large uncertainty on metallicities derived in this range. To break the degeneracy and provide tighter constraints in this regime a joint fit to multiple line ratios is required. This can be done using line ratios such as $R2$ and $O32$, where a calibration should be chosen that accounts for the ionisation parameter range of the inference sample.

\subsubsection{$\widehat{R}$ recalibration} \label{sec:rhat_recalibration}
In Figure \ref{fig:Rhat_calibration} we show the $\widehat{R}$ diagnostic in more detail. The left panel of this figure shows the systematic deviation of our metallicity measurements from the $\widehat{R}$ calibration by \cite{laseter2024}. In both our DESI EDR and EXCELS samples, the low-metallicity end of the $\widehat{R}$ diagnostic is shallower. We use the DESI EDR sample to refit the $\widehat{R}$ diagnostic, using orthogonal distance regression (ODR) as implemented in \textsc{Scipy} \citep{boggs1981, scipy2020}. We use a 3$^{\rm rd}$ order polynomial of the form $\widehat{R} = \sum c_i x^{i} $ where $x = 12 + \log({\rm O/H}) - 8.69$ to fit the data. The ODR fit is performed 500$\times$ using bootstrap resampling to estimate the uncertainties in the fitted parameters. The best fit coefficient values are $c_0 = 0.59_{-0.03}^{+0.03}$, $c_1 = -0.64_{-0.15}^{+0.13}$, $c_2 = -0.70_{-0.21}^{+0.19}$ and $c_3 = -0.10_{-0.10}^{+0.08}$. As our sample lacks coverage at the high-metallicity end, we only use our fitted relation where $\rm 12 + log(O/H) \leq 8.2$, and revert to the \cite{laseter2024} calibration at higher metallicities:
\begin{equation}
    \widehat{R}= 
\begin{cases}
    0.59 -0.64x -0.70x^2-0.10x^3, & \text{if } x\leq -0.49\\
    0.0492 -2.97x -3.97x^2-1.84x^3 -0.332x^4,               & \text{otherwise}
\end{cases}
\label{eq:rhat_this_work}
\end{equation}
In the right panels of Figure \ref{fig:Rhat_calibration} we show the residuals for the $\widehat{R}$ calibration in the $\Delta \widehat{R}$ and $\Delta$log(O/H) directions in the top and bottom panels, respectively. The 1-$\sigma$ residual scatter for the DESI EDR calibration sample is 0.06(0.08) dex in the $\Delta \widehat{R}$ direction and 0.14(0.13) dex in the $\Delta$log(O/H) direction for our calibration and the \cite{laseter2024} calibration in brackets. The residual scatter for the EXCELS sample, which was not used to calibrate the relation, is 0.07(0.09) dex in the $\Delta \widehat{R}$ direction and 0.11(0.14) dex in the $\Delta$log(O/H) direction, respectively. Notably, particularly at the low-metallicity end, our calibration performs significantly better for both the EXCELS and DESI EDR samples, as is shown in the panels on the right of Figure \ref{fig:Rhat_calibration}. The excellent agreement between local and high-redshift galaxies for the $\widehat{R}$ diagnostic is remarkable.

We also explore a non-parametric calibration similar to the method used by \cite{langeroodi2024}. Our main motivation for exploring a non-parametric calibration is to determine if there are additional dependencies which are difficult to capture using a parametric function. If such dependencies exist, the non-parametric calibration may provide superior metallicity estimates. We use the \textsc{SciPy} Gaussian kernel density estimation (KDE) implementation \citep[][]{silverman1986,scott1992,scipy2020}. We map the probability density distribution of the observational properties ($R2$, $R3$, EW(H$\beta$)) and metallicity of galaxies using a multivariate KDE. We then use the KDE to find the highest probability 12 + log(O/H) given the observed $R2$, $R3$ and EW(H$\beta$). We use the DESI EDR sample to calibrate the KDE and then apply the KDE method on both our EXCELS and DESI EDR samples to retrieve strong-line metallicities. The metallicities of the galaxies in the DESI EDR sample were derived using a ``leave-one-out'' approach. This means that a new KDE without a particular galaxy was created for each observation in the sample. This KDE was then used to infer the strong-line metallicity of that particular galaxy. The 1-$\sigma$ residual scatter for the DESI EDR calibration sample is 0.06 dex in the $\Delta \widehat{R}$ direction and 0.17 in the $\Delta$log(O/H) direction, respectively. The residual scatter for the EXCELS sample, which was not used to calibrate the relation, is 0.08 dex in the $\Delta \widehat{R}$ direction and 0.19 dex in the $\Delta$log(O/H) direction, respectively. The non-parametric approach performs well, yet does not provide a clear advantage over the parametric $\widehat{R}$ calibration. A larger calibration sample covering a wide range of metallicities may improve the performance of this approach as shown by the results from \cite{langeroodi2024}.

\begin{figure*}
	\begin{center}
	\includegraphics[width=\textwidth, trim=0 0 0 20.3cm, clip]{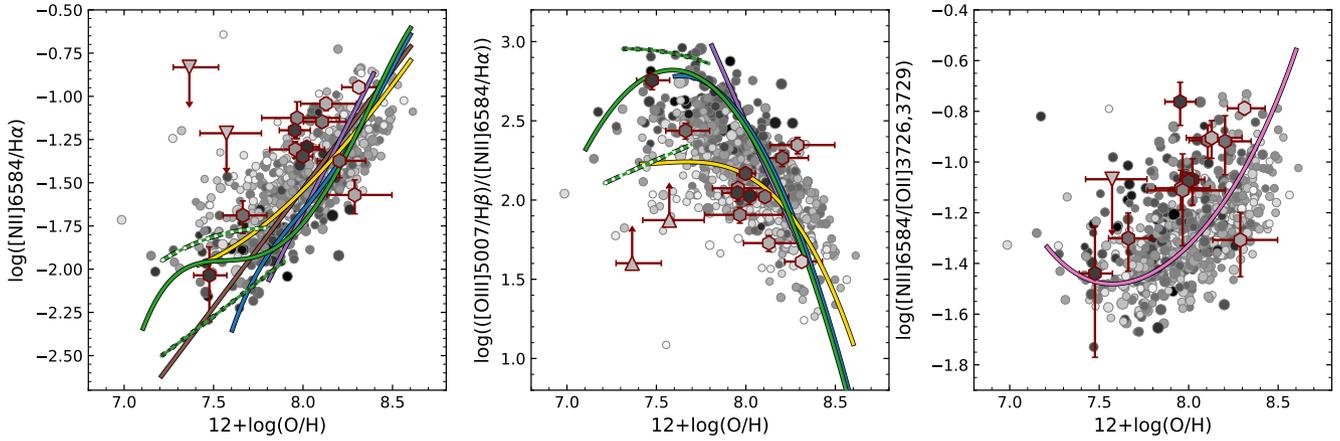}
	\caption{The relation between $T_e$-metallicity and nitrogen based strong-line ratios of the \textit{JWST} EXCELS (\textit{hexagons, maroon edges}) and the local sample of DESI EDR galaxies (\textit{circles, grey edges}). \textit{N2}, \textit{O3N2} and \textit{N2O2} (\textit{left-to-right} and \textit{top-to-bottom}) strong-line ratios are shown in the individual panels. The data points are coloured according to their \textrm{H}$\beta$ equivalent width as described by the colour bar in Figure \ref{fig:auroral_lineratios_hb_ew_oxygen}. Together with the measured data points we show several literature strong-line metallicity calibrations by \protect\cite{denicolo2002, kewley2002, bian2018, curti2020, sanders2021} and \protect\cite{nakajima2022} for which the colours are also described by the legend in Figure \ref{fig:auroral_lineratios_hb_ew_oxygen}. The \protect\cite{kewley2002} $N2O2$ calibration has been shifted by $-0.3$ dex in metallicity as this calibration was not derived using $T_e$-metallicities but photoionisation models. As we do not require the detection of the [\textsc{Nii}]$\lambda$6584 emission lines in our sample selection, we have included the 2-$\sigma$ upper limits for EXCELS galaxies where the wavelength of this emission line was observed but no line was detected (triangles with arrows). }
	\label{fig:auroral_lineratios_hb_ew_nitrogen}
	\end{center}
\end{figure*}

\begin{figure*}
	\begin{center}
	\includegraphics[width=\textwidth, trim=0 0 0 20.3cm, clip]{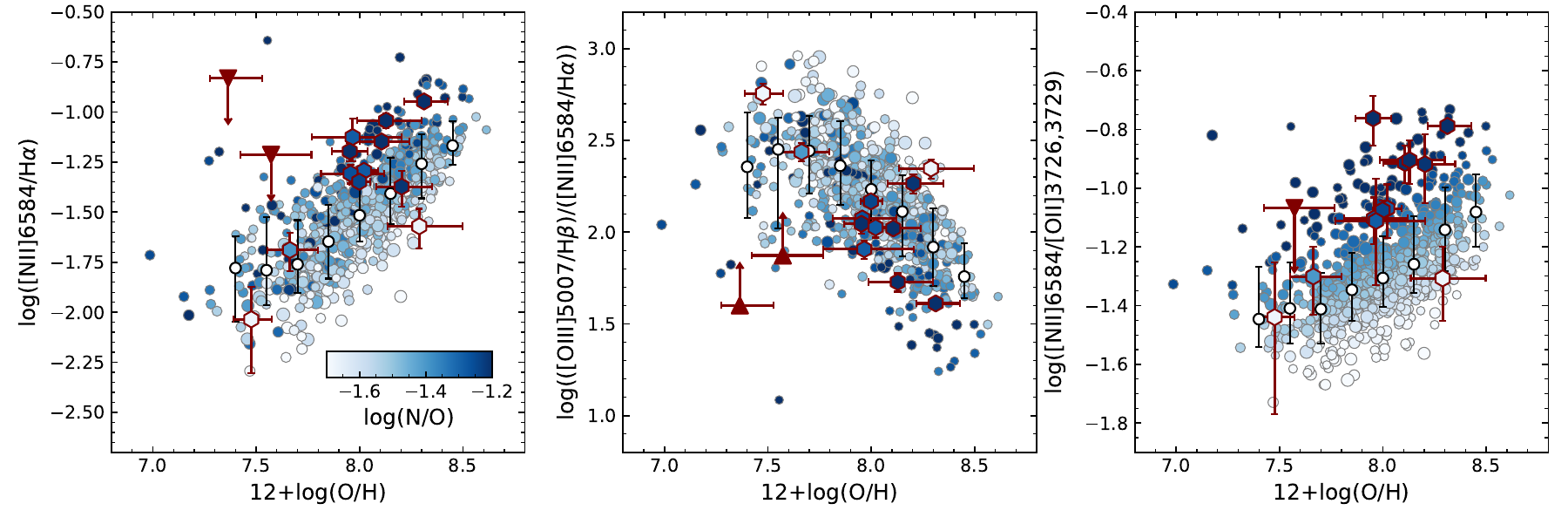}
	\caption{The same as Figure \ref{fig:auroral_lineratios_hb_ew_nitrogen} but with data points coloured using the nitrogen-to-oxygen abundance ratio as shown by the colour bar on the left panel. There are no reliable constraints on the N/O abundance ratio for the [\textsc{Nii}]$\lambda$6584 upper limit measurements, therefore these upper limits are coloured maroon and do not follow the colourbar shown in the left panel. In these panels we also include the median and 16$^{\rm th}$ to 84$^{\rm th}$ percentile DESI EDR line ratios at fixed O/H (\textit{black outlined circles with error bars}). See Figure \ref{fig:auroral_lineratios_hb_ew_oxygen} for full details of the samples shown.}
	\label{fig:auroral_lineratios_no_ratio_nitrogen}
	\end{center}
\end{figure*}

\subsection{Nitrogen line based calibrations}
The nitrogen [\textsc{Nii}]$\lambda$6584 emission line is also commonly used in metallicity diagnostics. An important common factor for all these nitrogen-based diagnostics is a sensitivity to variation in the nitrogen-to-oxygen abundance ratio (N/O). The average N/O abundance ratio is metallicity dependent and even at fixed metallicity there is a large scatter \citep[e.g.,][]{vanzee2006, izotov2006, perez-montero2009}. Furthermore, there is evidence for a systematic enhancement of nitrogen in galaxies (at fixed O/H) in the early Universe \citep[e.g.,][]{strom2017, strom2018, cameron2023, arellano-cordova2024excels, topping2024a, topping2024b}. Therefore, nitrogen-based emission line diagnostics derived at $z \sim 0$ can be systematically biased when applied in higher redshift regimes. In this work, we have information on the N/O abundance ratio for 714 out of the 782 DESI EDR galaxies and a subset of 12 of the 22 EXCELS galaxies where the [\textsc{Nii}]$\lambda$6584 line is detected and a further two where the [\textsc{Nii}]$\lambda$6584 line was observed but too faint for detection. 

Particularly, the monotonic trend of the $N2$ diagnostic is useful to break the degeneracies present in many of the oxygen line calibrations. The $O3N2$ and $N2O2$ diagnostics are also often used for metallicity measurements \citep[e.g.,][]{kewley2002,pettini2004}. The $O3N2$ diagnostic is a useful alternative for the $O32$ diagnostic when the [\textsc{Oii}]$\lambda\lambda$3726,3729 doublet is not within the observed wavelength range because the [\textsc{Nii}]$\lambda$6584 emission line has a similar ionisation potential. However, this means the $O3N2$ diagnostic does come with similar dependencies on ionisation parameter as the $O32$ diagnostic. The $N2O2$ diagnostic is insensitive to variations in the ionisation parameter; however, due to the large wavelength difference between the emission lines, this diagnostic does have a strong dependence on dust attenuation. 

In Figure \ref{fig:auroral_lineratios_hb_ew_nitrogen} we show the metallicity dependence of these nitrogen-based diagnostics. The residuals for each of the calibrations of these diagnostics are shown in more detail in Figure \ref{fig:auroral_lineratios_residuals} and \ref{fig:auroral_lineratios_metallicity_residuals} of Appendix \ref{sec:appendix_residuals_diagnostics}. The monotonic trend of the $N2$ diagnostic is clear in the DESI EDR sample; there is also consistent agreement in the literature on the general trend of the $N2$ calibration \citep[e.g.,][]{denicolo2002, bian2018, curti2020, sanders2021, nakajima2022}. Clear trends are also visible in the DESI EDR measurements of $O3N2$ and $N2O2$. However, the relation between the nitrogen strong-line ratios and metallicity is not as clear for the EXCELS sample. The EXCELS galaxies are scattered and many are offset from the local Universe trends in each panel. The dependence on nitrogen-to-oxygen abundance ratio may explain this offset. This is shown in Figure \ref{fig:auroral_lineratios_no_ratio_nitrogen}, which shows that many galaxies in the EXCELS sample have enhanced N/O abundance ratios compared to the average local population (as shown by the median DESI EDR line ratios at fixed metallicity, black outlined large circles with 16$^{\rm th}$ to 84$^{\rm th}$ percentile range error bars). In fact, we can see that the EXCELS galaxies are co-located with local galaxies with a similar N/O abundance ratio. These figures show the N/O dependence of nitrogen-based metallicity diagnostics and demonstrate that some caution should be used when using nitrogen-based empirical calibrations to determine oxygen abundances.

\begin{figure*}
\begin{center}
        \includegraphics[width=0.95\textwidth]{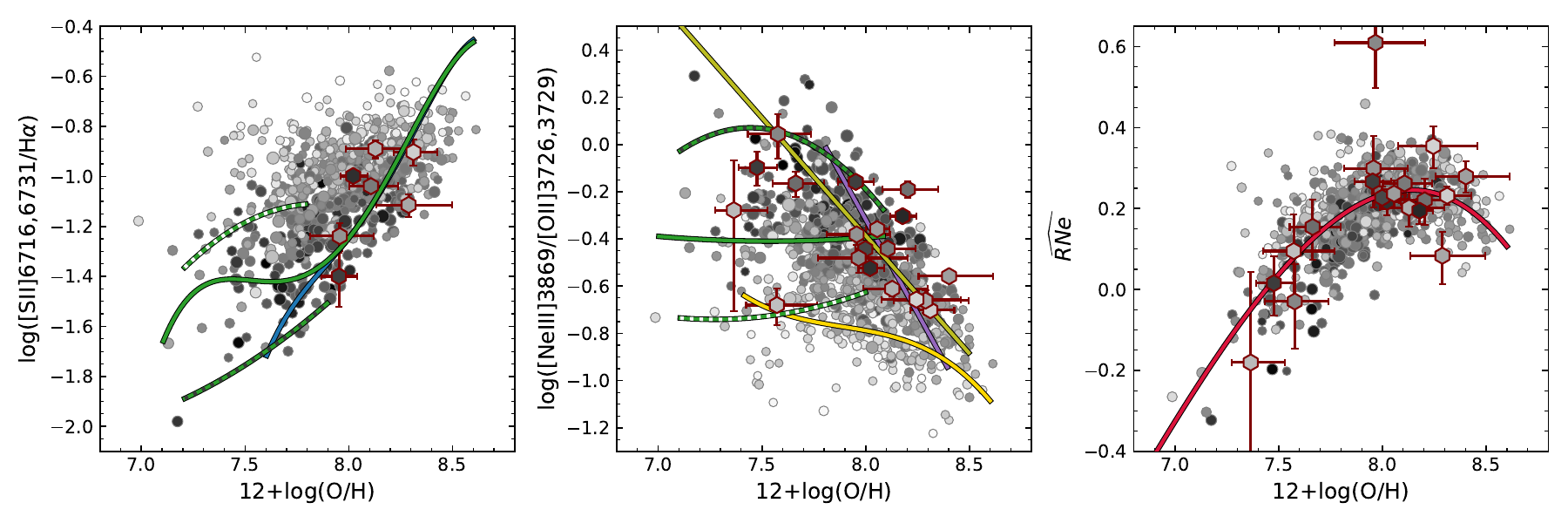}
        \caption{The relation between $T_e$-metallicity and the $S2$ (left), $Ne3O2$ (middle) and $\widehat{RNe}$ (right) strong-line ratios of the \textit{JWST} EXCELS (\textit{hexagons, maroon edges}) and the local sample of DESI EDR galaxies (\textit{circles, grey edges}). The data points are coloured according to their \textrm{H}$\beta$ equivalent width as described by the colour bar in figure \ref{fig:auroral_lineratios_hb_ew_oxygen}. Together with the measured data points we show several literature strong-line metallicity calibrations by \protect\cite{bian2018, curti2020, sanders2021, nakajima2022} and \protect\cite{sanders2024} for which the colours are also described by the legend in Figure \ref{fig:auroral_lineratios_hb_ew_oxygen}.}
        \label{fig:auroral_lineratios_hb_ew_neon_sulfur}
\end{center}
\end{figure*}

\begin{figure*}
	\begin{center}
	\includegraphics[width=\textwidth]{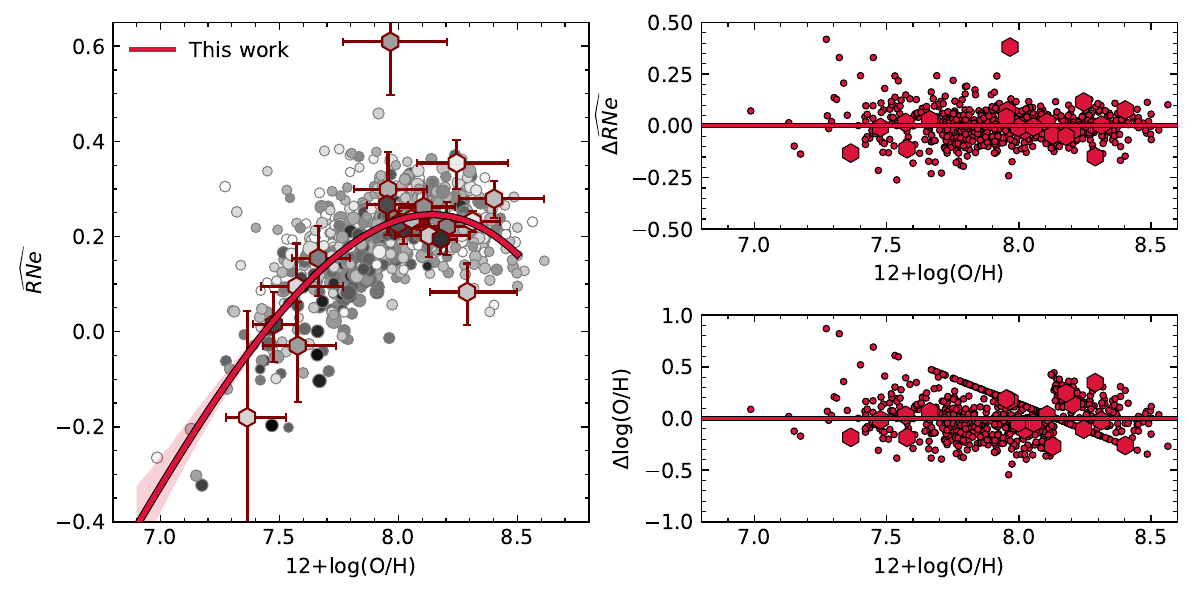}
	\caption{\textit{Left}: An enlarged version of the $\widehat{RNe}$ panel in Figure \ref{fig:auroral_lineratios_hb_ew_neon_sulfur} where the data points are coloured using the colourbar in Figure \ref{fig:auroral_lineratios_hb_ew_oxygen}. We also show our calibration based on the DESI EDR data (Eq. \ref{eq:nehat_this_work}, red line) along with the 1-$\sigma$ fit uncertainties shaded in light pink. \textit{Top-right}: The residuals in the $\Delta \widehat{RNe}$ direction. \textit{Bottom-right}: Same as top-right, however, here showing residuals in the $\Delta$log(O/H) direction. See Section \ref{sec:rnehat_calibration} and Appendix \ref{sec:appendix_residuals_diagnostics} for further details on the residuals of the calibration.}
	\label{fig:RNehat_gamma_calibration}
	\end{center}
\end{figure*}

\subsection{Sulfur and neon line calibrations} \label{sec:neon_sulfur_calibrators}
Sulfur and neon are both alpha-elements, which means that, unlike nitrogen, their abundances evolve in tandem with oxygen \citep[confirmed up to $z=5.3$ by e.g.,][]{henry1989,garnett2002, stanton2024excels, arellano-cordova2024classy, esteban2025}. Therefore, there is less concern regarding evolving abundance ratios than with nitrogen, both as a function of metallicity and through systematic redshift evolution. Emission lines of these elements are routinely observed in the \textit{JWST} spectra of high-redshift galaxies. Hence, their use in metallicity calibrations should be considered.

We use the $S2$ diagnostic to study how well we can recover metallicities using sulfur emission lines. In the right panel of Figure \ref{fig:auroral_lineratios_hb_ew_neon_sulfur} we show the metallicity dependence of the $S2$ diagnostic. This figure displays only a subsample of 7 of the galaxies in our EXCELS sample due to the long wavelength of the [\textsc{Sii}]$\lambda\lambda$6716,6731 emission lines which are redshifted out of wavelength coverage at $z\sim6.8$ in the G395/F290LP filter and already at $z\sim3.7$ in the G235M/F170LP filter. Besides the small number of EXCELS galaxies, the $S2$ diagnostic yields very similar results to the $R2$ diagnostic with a strong Pearson correlation coefficient of 0.92 between the two diagnostics. Therefore, the behaviour and caveats applicable to the $R2$ diagnostic are applicable here as well.

The [Ne\textsc{iii}]$\lambda$3869 emission line is potentially much more useful for the study of high-redshift galaxies due to its location at the blue end of the rest-frame optical. We detect the [Ne\textsc{iii}]$\lambda$3869 emission line in 19 out of our 22 EXCELS targets. For the remaining 3 targets the [Ne\textsc{iii}]$\lambda$3869 emission line was located in a detector gap of the observed spectrum. The $Ne3O2$ diagnostic (left panel of Figure \ref{fig:auroral_lineratios_hb_ew_neon_sulfur}) is observable to $z\sim12.7$ with \textit{JWST}/NIRSpec. This diagnostic also benefits from the close proximity in wavelength of all its emission lines, which makes the $Ne3O2$ ratio insensitive to dust attenuation and observable with minimal spectral coverage. There is a general trend between metallicity and $Ne3O2$, with a large EW(H$\beta$) dependent scatter which is well captured by the \citep[][]{nakajima2022} calibrations. Most of our EXCELS galaxies follow the calibrations by \cite{bian2018} and \cite{sanders2024}, with some clear exceptions for low EW(H$\beta$) galaxies due to the ionisation parameter dependence. The similar behaviour of the $Ne3O2$ and $O32$ diagnostics is due to similar ionisation potentials which result in a very strong linear correlation between the relative line strengths of [Ne\textsc{iii}]$\lambda$3869 and [\textsc{Oiii}]$\lambda$5007. For the DESI EDR sample, the Pearson correlation coefficient is 0.85 between the $R3$ and $Ne3$ diagnostics \citep[see also][]{reddy2023}. This means that similar behaviour can be expected between any diagnostics in which the [\textsc{Oiii}]$\lambda$5007 line is replaced by [Ne\textsc{iii}]$\lambda$3869 (assuming that the typical Ne/O abundance ratio does not evolve with redshift). 

\subsubsection{$\widehat{RNe}$ calibration} \label{sec:rnehat_calibration}
We propose a new line diagnostic $\widehat{RNe}$; this is a neon-based version of the $\widehat{R}$ diagnostic \citep{laseter2024} which minimises the dependence on ionisation parameter. The diagnostic is defined as

\begin{equation}
\begin{split}
    \widehat{RNe} &= 0.47 \times \log \left[ \frac{[\textsc{Oii}]\lambda\lambda3726,3729}{\textsc{H}\gamma} \right]\\ 
     & \quad + 0.88 \times \log \left[ \frac{[\textrm{Ne}\textsc{iii}]\lambda3869}{\textsc{H}\gamma} \right]. \\
\end{split}
\label{eq:nehat_diagnostic}
\end{equation}
We show this in the right panel of Figure \ref{fig:auroral_lineratios_hb_ew_neon_sulfur} and 
in Figure \ref{fig:RNehat_gamma_calibration}. Due to the very strong linear correlation between $R3$ and $Ne3$ we do not need to separately find the optimum combination of line ratios needed to reduce the ionisation parameter dependence; instead, we can simply use the values found by \cite{laseter2024} for the $\widehat{R}$-diagnostic. This new line diagnostic shows a clear correlation with metallicity, albeit with slightly larger scatter than $\widehat{R}$ due to the relative faintness of [Ne\textsc{iii}]$\lambda$3869 and H$\gamma$ compared to [\textsc{Oiii}]$\lambda$5007, and H$\beta$. However, in addition to the insensitivity to ionisation parameter, the $\widehat{RNe}$ diagnostic has two other advantages. The diagnostic is applicable to very high-redshift \textit{JWST}/NIRSpec observations up to $z\sim11.2$; this is particularly useful in the wavelength range between $9.5<z<11.2$ where the emission lines required for the $\widehat{R}$ diagnostic are no longer observable. The $\widehat{RNe}$ line ratio is also mostly insensitive to dust attenuation due to the close wavelength proximity of all the required emission lines.

We perform a similar line fit to this relation as we did for the $\widehat{R}$ diagnostic. The ODR fitting is done 500$\times$ using bootstrap resampling to estimate the uncertainties in the fitted parameters. We use a 3$^{\rm rd}$ order polynomial of the form $\widehat{RNe} = \sum c_i x^{i} $ where $x = 12 + \log({\rm O/H}) - 8.69$ to fit the data. The best fit coefficient values are $c_0 = 0.04_{-0.04}^{+0.03}$, $c_1 = -0.78_{-0.17}^{+0.13}$, $c_2 = -0.82_{-0.22}^{+0.18}$ and $c_3 = -0.14_{-0.09}^{+0.08}$. Therefore the function describing our $\widehat{RNe}$ calibration is
\begin{equation}
    \widehat{RNe}= 0.04 -0.78x -0.82x^2-0.14x^3.
\label{eq:nehat_this_work}
\end{equation}
We show the fitted relation and residuals along the $\Delta \widehat{RNe}$ and $\Delta$log(O/H) directions in Figure \ref{fig:RNehat_gamma_calibration}. The 1-$\sigma$ residual scatter for the DESI EDR calibration sample is 0.07 dex and 0.18 dex along $\Delta \widehat{RNe}$ and $\Delta$log(O/H), respectively. The residual scatter for the EXCELS sample which was not used in the fitting of the line ratio is 0.11 dex and 0.16 dex, respectively. This shows that our $\widehat{RNe}$ calibration performs excellently for both local and high-redshift galaxies. Similar to  the $\widehat{R}$ diagnostic there is a region at $\widehat{RNe}\gtrsim0.15$ where this diagnostic is largely insensitive to metallicities between $7.8 \lesssim 12+\rm log(O/H) \lesssim 8.3$; due to this there is a large uncertainty on metallicities derived in this range. The $Ne3O2$ diagnostic, which uses the same emission lines, can be used to break the degeneracy in the plateau regime. Here a calibration should be chosen that matches the ionisation parameter of the inference sample. At high-redshift the \cite{sanders2024} calibration could be used for this purpose.

\subsection{Summary on strong-line calibrations}
In general, we find that many strong-line diagnostics are heavily dependent on parameters other than metallicity. Most notable is the strong dependence on the ionisation parameter of e.g.,{} $R2$, $R3$, $O32$, $O3N2$, $S2$ and $Ne3O2$ and the N/O abundance ratio dependence of nitrogen-based calibrations. Line calibrations using neon and sulfur emission lines do not suffer from abundance ratio biases due to systematic deviations of their abundances relative to oxygen. This is because the abundances of these alpha-elements evolve in tandem. Due to the systematic redshift evolution of both the average ionisation conditions (i.e., increasing ionisation parameter and harder ionising spectra) in star-forming regions and N/O abundance ratios at fixed metallicity, this means that strong-line metallicity calibrations using these line ratios are systematically biased if the ionisation parameters or N/O abundance ratios of the sample for which metallicities are inferred are different from the calibration sample. This can be addressed by accounting for these secondary dependencies through explicit inclusion of the dependence in line calibrations or by matching the properties of the calibration and inference samples. We also show that there are line calibrations such as $\widehat{R}$ and $\widehat{RNe}$ which are robust against the evolution of ionisation conditions. We provide our own calibrations for these line ratios based on the local DESI EDR sample in Equations \ref{eq:rhat_this_work} and \ref{eq:nehat_this_work}, respectively.

\begin{figure*}
    \begin{center}
        \includegraphics[width=0.7425\textwidth]{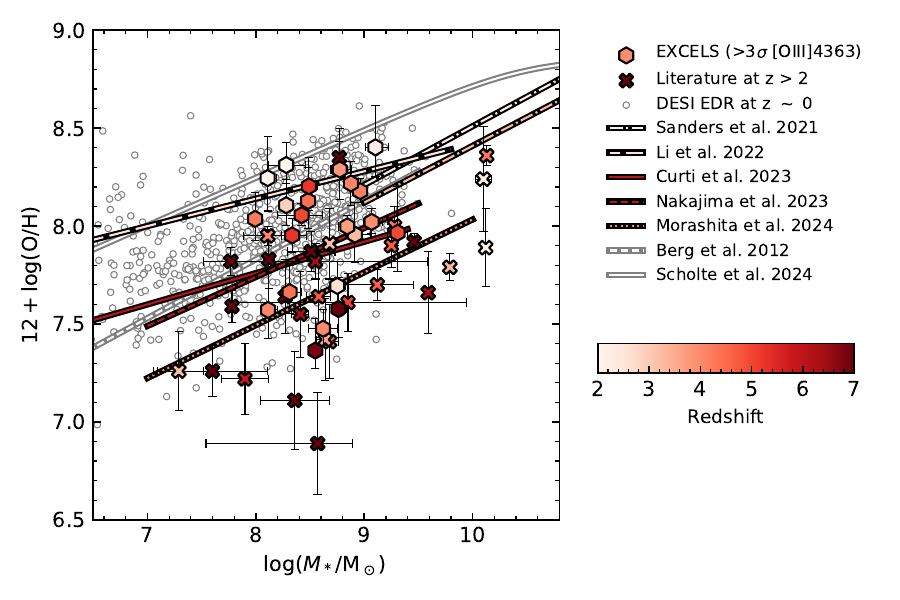}
    \caption{[\textsc{Oiii}]$\lambda$4363 detected galaxies on the mass-metallicity plane. The mass-metallicity relation for our EXCELS measurements (black outlined hexagons and diamonds), DESI measurements (grey outlined circles) and $z>2$ literature values \protect\citep[crosses;][]{sanders2023metallicity, nakajima2023, curti2023, morishita2024, cullen2025}. Each datapoint is coloured according to the redshift which is shown by the colourbar. Several local and high-redshift parametrisations of the MZR are shown as described in the legend.}
    \label{fig:mzr}
    \end{center}
\end{figure*}

\begin{figure*}
    \begin{center}
    \begin{subfigure}[b]{0.495\textwidth}
        \includegraphics[width=\textwidth]{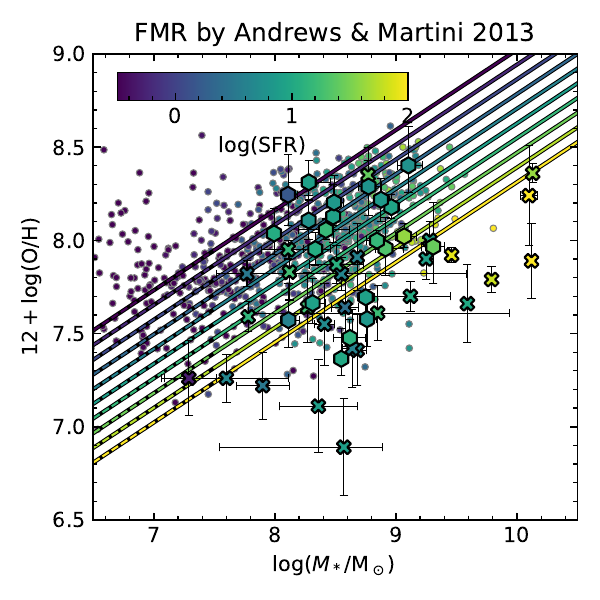}
    \end{subfigure}
    \hfill
    \begin{subfigure}[b]{0.495\textwidth}
        \includegraphics[width=\textwidth]{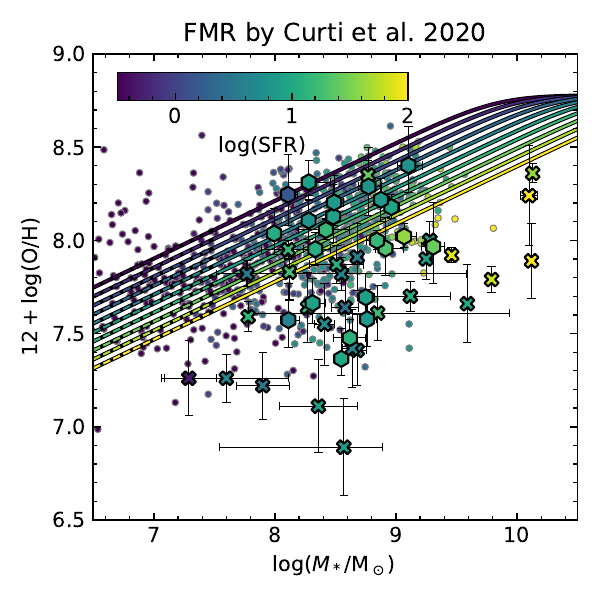}
    \end{subfigure}
    \caption{Our sample of EXCELS galaxies (hexagons and diamonds) on the mass-metallicity plane coloured by star formation rate, shown together with $z>2$ literature values \protect\citep[crosses;][]{sanders2023metallicity, nakajima2023, curti2023, morishita2024, cullen2025} and DESI EDR local galaxies (small circles). We show the fundamental metallicity relation by \protect\cite{andrews2013} and \protect\cite{curti2020} in black outlined lines in the left and right panels, respectively. These lines are also coloured by the star formation rate.}
    \label{fig:fmr}
    \end{center}
\end{figure*}

\begin{figure}
    \begin{center}
    \begin{subfigure}[b]{0.495\textwidth}
        \includegraphics[width=\textwidth]{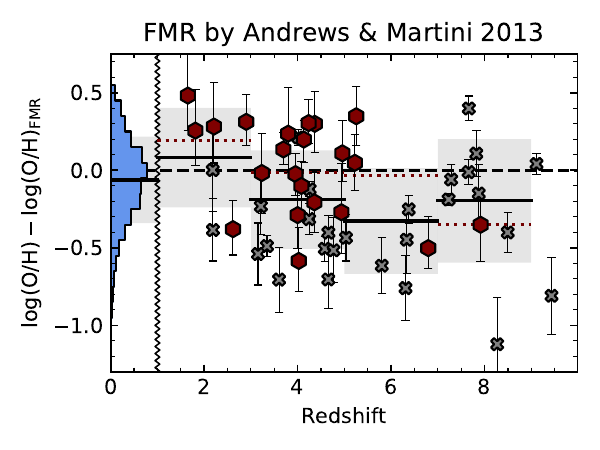}
    \end{subfigure}
    \hfill
    \begin{subfigure}[b]{0.495\textwidth}
        \includegraphics[width=\textwidth]{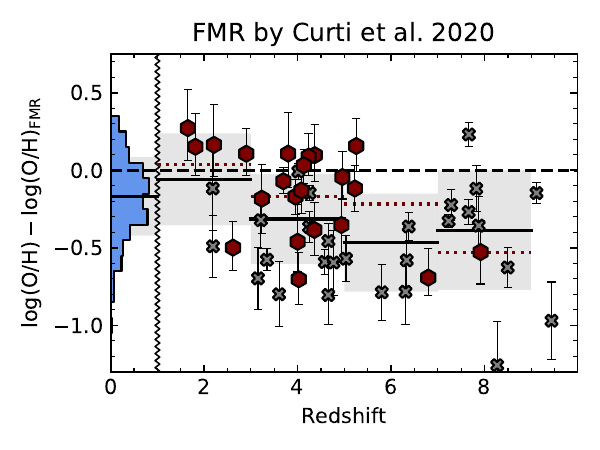}
    \end{subfigure}
	\caption{The redshift dependence of the residuals of the EXCELS (hexagons and diamonds) and literature (crosses) measurements with respect to the \protect\cite{andrews2013} parametrisation of the FMR in the top panel and the \protect\cite{curti2020} FMR in the bottom panel. We also show the residuals distribution of the DESI EDR local sample in the blue histograms. The local sample and higher redshift measurements are separated by a black zig-zag line. Binned averages and standard deviations of the residuals are shown in black lines and grey bands for the full sample of our measurements and the literature, respectively. The binned averages of our EXCELS measurements are shown in maroon dotted lines.}
	\label{fig:delta_fmr}
	\end{center}
\end{figure}

\section{Location on the mass-metallicity relation and fundamental metallicity relation} \label{sec:mzr}
In this section we will use our measurements to understand the relation between the chemical enrichment of these galaxies and their overall evolution. We look at the location of galaxies with $T_e$-metallicities on the mass-metallicity relation and fundamental metallicity relation. We show our galaxies on the MZR in Figure \ref{fig:mzr}. Our sample of galaxies resides systematically below the average MZR in the local Universe; there are two main reasons for this: (1) the bias of the sample of galaxies studied due to strict selection criteria and (2) the redshift dependent evolution of the MZR \citep[e.g.,][]{savaglio2005, moustakas2011, zahid2013, cullen2014, juneau2014, sanders2021, stanton2024kmos}. The bias introduced by the requirement of the [\textsc{Oiii}]$\lambda$4363 detection is such that by selection $\rm 12+log(O/H) \lesssim 8.5$. Additionally, we are also biased towards highly star-forming galaxies which display the brightest emission lines. This is due to the fact that at higher metallicities and lower star formation rates the [\textsc{Oiii}]$\lambda$4363 emission line becomes too faint to be detected. This upper limit is similar for both our EXCELS and DESI samples. We are unable to make any robust inference of the MZR of the full galaxy population from this subset. As a general trend we do observe that the galaxies in our sample are increasingly metal-poor as a function of redshift. This is in general agreement with the redshift evolution of the MZR. We can see that our results are broadly in agreement with the measurements of $z>2$ [\textsc{Oiii}]$\lambda$4363 detected galaxies in the literature \citep[see Section \ref{sec:literature_sample}, data taken from][]{sanders2023metallicity,curti2023,nakajima2023,morishita2024,cullen2025}. Similar to many other studies, we also find that the lowest metallicity galaxies are at $\rm 12+log(O/H)\sim7.2$ (0.032 Z$_{\odot}$). To date, very few galaxies have been found where $\rm 12+log(O/H)<7.0$ \citep[but see e.g.,][]{cullen2025}.

Whilst it is difficult to disentangle biases from population evolution on the mass-metallicity relation, this is less of a problem for the fundamental metallicity relation. This is because the FMR predicts that individual galaxies reside on a redshift independent 2D surface spanned by the 3D parameter space of stellar mass, metallicity and star formation rate \citep{mannucci2010}. As the $T_e$-metallicity measurements are some of the most reliable metallicities, our EXCELS measurements together with the literature sample described in Section \ref{sec:literature_sample} \citep[][]{sanders2023metallicity,curti2023,nakajima2023,morishita2024,cullen2025} is an excellent sample to verify whether these galaxies are in agreement with the fundamental metallicity relation. Particularly at the highest redshifts, where the number of EXCELS galaxies is low, the contribution of the compiled literature is a valuable addition. There are many different measurements of the FMR; however, here we focus on the \cite{andrews2013} and \cite{curti2020} calibrations, which are both rooted in $T_e$-metallicity measurements. 

In Figure \ref{fig:fmr} we show the FMR derived by \cite{andrews2013} in the left panel and \cite{curti2020} in the right panel. It is clear from this figure that many of the EXCELS and $z>2$ literature galaxies do not align closely to the $z=0$ FMR. This can be seen by the large number of EXCELS and $z>2$ literature galaxies below the FMR relation with the lowest metallicities (and highest SFRs) shown. This is most prominently the case for the \cite{curti2020} FMR parametrisation. The redshift dependence of the residuals between the metallicity of galaxies and their expected metallicity based on the FMR is shown in Figure \ref{fig:delta_fmr}. This figure shows that there is a large scatter in the residuals of the FMR of high-redshift galaxies. There are a large number of high-redshift galaxies with metallicities below the FMR expectation. However, the systematic trends through the redshift binned average residuals are suggestive but do not show a statistically significant deviation from the FMR (maximum deviation at $\sim1.3\sigma$ for \cite{curti2020} parametrisation at $5<z<7$). We do note that the trends seen here are not dissimilar from  results by e.g., \cite{heintz2023}, \cite{curti2023}, \citep[][]{nakajima2023}, who found a deviation from the FMR for galaxies at $z>4$; their results are based on larger samples of galaxies using strong-line metallicity measurements.

There is also significant scatter for the local DESI EDR sample, and a mild deviation from the \cite{curti2020} FMR-parametrisation. This deviation is very similar to the offset of local blueberry and green pea galaxies \citep{yang2017}. We note that due to the small aperture of the fibers of the DESI instrument (diameter $\sim 1.5$ arcsec) the observed spectra only capture the emission from the central regions of galaxies. Therefore, the derived metallicities are representative of these central regions. This may introduce some biases towards higher metallicity in central regions through galactic metallicity gradients (opposite to the direction of the deviation from the FMR). We also note that star formation rate measurements for the DESI sample are derived using SED fitting.

The significance of the deviation from the FMR is low on a population basis, however, on an individual galaxy basis there are much more significant deviations from the FMR. Several of our EXCELS and $z>2$ literature galaxies have a larger than $5\sigma$ deviation from the local FMR. This is true for both parametrisations of the FMR, however, the number of $5\sigma$ deviations is larger for the \cite{curti2020} parametrisation. Significant deviations from the FMR are fairly common at higher redshifts, however, rare in our local DESI EDR sample. The significance of these offsets from the FMR is dependent on the reliability of uncertainties on stellar mass, star formation rate and metallicity as well as an assumption of a Gaussian probability distribution for these parameters. An underestimate of these uncertainties could inflate the true level of discrepancy with the FMR.

\section{Discussion} \label{sec:discussion}
We have presented a sample of 22 high-redshift [\textsc{Oiii}]$\lambda$4363 detected galaxies which were observed as part of the \textit{JWST}-EXCELS survey \citep{carnall2024}. We determined their chemical abundances using the $T_e$-method; these measurements highlight the fast-growing sample of galaxies in the early Universe for which we can robustly infer chemical abundances \citep[see also e.g.,][]{arellano-cordova2022, sanders2023metallicity,stanton2024excels, arellano-cordova2024excels}. Our measurements add to other recent samples of high-redshift [\textsc{Oiii}]$\lambda$4363 detections such as \cite{curti2023, nakajima2023, sanders2024, laseter2024} and \cite{morishita2024} and highlight the value of deep spectroscopy of large numbers of high-redshift galaxies. Simultaneously, we have presented our metallicity measurements for a high-quality sample of spectra taken from the DESI Early Data Release \citep{desi2023edr}. These low-redshift spectra probe a large sample of very faint and low-mass galaxies which had not been targeted by previous generations of large spectroscopic surveys such as SDSS \citep{york2000}. In the following sections, we discuss the implications of our measurements for strong-line metallicity calibrations and chemical enrichment over cosmic time.

\subsection{Strong-line metallicity calibrations}
We have presented the metallicity dependence of a large number of commonly used strong-line diagnostics based on the rest-frame optical emission lines of hydrogen, oxygen, nitrogen, neon and sulfur. We compared our high-redshift EXCELS and local DESI EDR $T_e$-metallicity measurements to the $T_e$-derived strong-line calibrations of \cite{denicolo2002, kewley2002, bian2018, curti2020, sanders2021,nakajima2022,sanders2024} and \cite{laseter2024}. The comparisons show that for many strong-line diagnostics there is a strong secondary dependence on ionisation parameter. However, the $\widehat{R}$ and the new $\widehat{RNe}$ diagnostics are insensitive to ionisation parameter and therefore can be applied at any redshift. Finally, we find that nitrogen based strong-line diagnostics may be biased at high-redshift due to moderately enhanced N/O abundance ratios.

\subsubsection{Dependence on ionisation parameter}
The main driver of the observed scatter in many of these strong-line diagnostics is clear: they are heavily dependent on the ionisation parameter. This dependence has been known for several decades \citep[e.g.,][]{mcgaugh1991,pilyugin2000,pilyugin2001, kewley2002,pilyugin2005, blanc2015,nakajima2022} and should be addressed when using strong-line metallicity measurements. This is of particular relevance when deriving metallicities for high-redshift galaxies, which have been shown to have increased ionisation parameters, alpha-enhanced abundance ratios (i.e.{}, iron deficit relative to oxygen) and, as a result, systematically harder ionising spectra at fixed metallicity \citep[e.g.,][]{steidel2014,shapley2015,sanders2016,kashino2017,strom2017,strom2018,kashino2022,sanders2023excitation,shapley2024, chartab2024, stanton2024kmos, stanton2024excels}. If this dependence is not addressed, this can introduce biases and artificial metallicity trends with other galaxy properties (e.g., star formation rate). Large enough samples of galaxies with $T_e$-metallicities are emerging such that secondary dependencies can be addressed reliably. There are many methods that explicitly address this secondary dependence \citep[e.g.,][]{pilyugin2000, pilyugin2001, kewley2002,kobulnicky2004, pilyugin2005, blanc2015, nakajima2022, marconi2024, laseter2024, langeroodi2024, chakraborty2024}. Others carefully match the properties of calibration samples to the samples for which metallicities are inferred \citep[e.g.,][]{bian2018, sanders2024}. Strongly ionisation parameter dependent diagnostics include $R2$, $R3$, $O32$, $O3N2$, $Ne3O2$ and $S2$. 

We presented new $T_e$-metallicity measurements from two new datasets: the \textit{JWST}/EXCELS survey and the Early Data Release of the DESI survey. These new data are examples of valuable new data which will allow us to improve strong-line metallicity calibrations, particularly to be applied to observations at high-redshift. The dataset from the DESI EDR presented here and the coming data from the DESI survey will allow us to derive much more fine-grained strong-line calibrations that include secondary dependencies in the near future (X. Tan et al. in prep). The large amount of DESI data becoming available will make it possible to expand on successful strong-line calibration methods which account for ionisation parameter dependence \citep[e.g.][]{nakajima2023}. Simultaneously, the increasing samples of high-redshift galaxies with auroral line detections will allow us to robustly validate strong-line calibrations against the systematic redshift evolution of the properties of star-forming regions. Another method which may benefit from the growing local samples of auroral line detected galaxies are calibrations using local analogue samples. In this approach, the sample used to construct a strong line calibration has to be carefully selected such that it matches the ionisation parameter distribution of the sample on which the calibration is applied. An example of this is the strong line calibrations using local analogues by \citep[][]{bian2018}. Alternatively, strong-line calibrations can now also be derived in-situ through the emerging samples of high-redshift $T_e$-metallicities such as the EXCELS sample presented here \citep[e.g.,][]{sanders2024}. Lastly, non-parametric methods such as the approach by \cite{langeroodi2024} will also benefit from these growing samples.

We find a strong and redshift-invariant metallicity dependence of the $\widehat{R}$ diagnostic with our local DESI EDR and high-redshift EXCELS sample. We do, however, see a shallower trend than the original \cite{laseter2024} calibration at the lowest metallicities in both our samples. Therefore, we provide our own calibration of this diagnostic. Further work is needed to fully understand the nature of the discrepancy between our measurements and those by \cite{laseter2024}. Discrepancies may be due to differences in our methodologies or sample selection. It is worth noting that the sample of galaxies targeted by DESI has significantly fainter magnitude limits than e.g.,{} SDSS Main Galaxy Survey \citep[][]{york2000,desi2023, desi2023edr}. This may indicate that the scatter is larger than one would naively assume from the observed samples (this statement may be true for all metallicity diagnostics). Analysis using theoretical photoionisation models may aid in improving our understanding of the observed $\widehat{R}$-diagnostic. Whilst the $\widehat{R}$ is a promising oxygen-based strong-line calibration, there is still a region at $\widehat{R}\gtrsim0.7$ where this diagnostic is largely insensitive to metallicities between $7.8 \lesssim 12+\rm log(O/H) \lesssim 8.3$; due to this there is a large uncertainty on metallicities derived in this range.

A recent paper by \cite{chakraborty2024} showed that, based on a calibration sample of high-redshift galaxies, a more narrow sequence than the \cite{laseter2024} $\widehat{R}$ can be achieved by using a different relative combination of $R2$ and $R3$: $\widehat{R}_{\rm Chakraborty} = 0.18\times R2 + 0.98 \times R3$. As shown in Appendix \ref{sec:rhat_chakraborty}, this alternative projection does not provide an improvement over the $\widehat{R}$ projection by \cite{laseter2024} for our DESI and EXCELS galaxies. An increased ionisation parameter dependence of this new projection is visible in Figure \ref{fig:rhat_chakraborty} compared to the left panel of Figure \ref{fig:Rhat_calibration}. However, the differences between the $R2$, $R3$ and $12+\log(\rm O/H)$ relations of these different samples may hold information on the systematic discrepancies between the $\widehat{R}$-calibrations of \cite{laseter2024}, \cite{chakraborty2024} and this work. 

In this work we introduced a new ionisation parameter independent line diagnostic, $\widehat{RNe}$, as shown in Equations \ref{eq:nehat_diagnostic} and \ref{eq:nehat_this_work}. This new diagnostic shows a very clear metallicity dependence (see Fig. \ref{fig:RNehat_gamma_calibration}) and does not show any systematic evolution with redshift. The main advantages of this new diagnostic are: (1) the lack of dependence on ionisation parameter, (2) insensitivity to dust attenuation due to the close wavelength proximity of all required emission lines, and (3) due to the location of all required lines in the blue end of the rest-frame optical this line ratio can be used at very high redshifts e.g., using \textit{JWST}/NIRSpec at $z\sim11.2$. Therefore, this diagnostic  allows us to measure the metallicities of galaxies from the present-day ($z=0$) to the era where some of the first galaxies are formed in the early Universe ($z\sim11$).

\subsubsection{Dependence on the N/O abundance ratio}
Metallicity calibrations using the [\textsc{Nii}]$\lambda$6584 emission line are commonly used for galaxies in the nearby Universe. In particular, the monotonic behaviour of the $N2$ calibrator can be valuable to determine metallicities at $12+\textrm{log(O/H)} \sim 8.0$, where many oxygen line calibrations reach a plateau. As \textit{JWST} now makes it possible to observe the [\textsc{Nii}]$\lambda$6584 line to $z\sim 7.0$, we investigated whether these calibrations can be applied in the high-redshift Universe. We find that the majority of our EXCELS measurements are offset from the average local nitrogen-based strong-line diagnostics (see Figure \ref{fig:auroral_lineratios_hb_ew_nitrogen}). However, we find that once the N/O abundance ratio is taken into account, the high-redshift galaxies inhabit a similar location on these diagrams as local galaxies with similar N/O abundance ratios (see Figure \ref{fig:auroral_lineratios_no_ratio_nitrogen}). The offset is due to the fact that our high-redshift galaxies are moderately more nitrogen-rich than the average local galaxy. This result is consistent with \cite{strom2018}, who found that $z\sim2$ galaxies that are offset from the $z=0$ star-forming sequence on the [\textsc{Nii}]-BPT diagram are nitrogen-rich compared to galaxies more closely resembling $z=0$ galaxies in the [\textsc{Nii}]-BPT diagram. It is also in agreement with more recent results of systematic enhancement of the N/O abundance ratio of high-redshift galaxies \citep[e.g.,][]{cameron2023, arellano-cordova2024excels, topping2024a, topping2024b}. The sensitivity of nitrogen based strong-line calibrations to the N/O abundance ratio was also found by \cite{arellano-cordova2020} for Galactic H\textsc{ii}-regions. Moreover, \cite{patricio2018} found that nitrogen strong-line calibrations systematically overestimate the metallicity of galaxies at redshifts between $1.4<z<3.6$.

A moderately enhanced N/O abundance ratio is expected as, at fixed $\rm log(O/H)$, high-redshift galaxies are typically more massive than $z=0$ galaxies. This implies that there is a larger contribution to the chemical abundances from evolved lower-mass stars in high-redshift galaxies while their $\rm log(O/H)$ remains low due to pristine inflows of gas \citep[e.g.,][]{recchi2008,amorin2010,lilly2013}. Additionally, it is possible that there are other processes which further enhance the nitrogen content of early galaxies, such as the presence of very massive stars or globular cluster formation \citep[e.g.,][]{vink2024, marques-chaves2024}. Systematically biased N/O abundance ratios directly impact the nitrogen-based strong-line calibrations. We conclude that some caution should be used when using nitrogen line calibrations to derive oxygen abundances, particularly at high-redshift.

\subsection{Chemical enrichment over cosmic time}
Improving our understanding of the chemical enrichment process of our Universe over cosmic time is one of the main reasons why we should aim to derive robust metallicity measurements of large samples of galaxies spanning a wide range of redshifts (EXCELS $1.65<z<7.92$, DESI EDR $z<0.48$). Our EXCELS $T_e$-metallicity measurements contribute to this aim as they provide a significant sample of robust metallicity measurements over a wide range of redshifts. We combined our measurements with other recent literature $T_e$-measurements to study the level of chemical enrichment of galaxies as a function of cosmic time and to understand whether the equilibrium between gas flows, star formation and chemical enrichment that gives rise to the FMR in the local Universe also applies to high-redshift galaxies \citep[e.g.,][]{mannucci2010,lilly2013}. 

Several studies have now shown that galaxies may systematically deviate from the FMR at high-redshift \citep[e.g.,][]{nakajima2023,heintz2023, curti2024}. This was shown by \cite{curti2024} at $z\gtrsim$; their study includes 146 galaxies at $3<z<10$ which were observed as part of the JADES programme. \cite{heintz2023} found a similar deviation at $z>7$ for a sample of 16 very high-redshift galaxies. A similar investigation by \cite{nakajima2023} based on 135 galaxies observed in the ERO \citep[SMACS 0723 lensing cluster;][]{pontoppidan2022}, GLASS and CEERS programmes concluded that a systematic deviation from the FMR is indeed present when the parametrisation by \cite{curti2020} is used, however, the deviation does not exist for the \cite{andrews2013} parametrisation. These studies are based on strong-line metallicity measurements derived using different, but similar metallicity calibrations as they have a large overlap in their calibration sample (the \cite{curti2020} and \cite{nakajima2022} calibrations, respectively). An important takeaway from these results is that there are two clear uncertain factors in understanding whether there is redshift evolution of the fundamental metallicity relation: (1) the calibration of the FMR and (2) the reliability and potential biases in strong-line metallicity calibrations. A homogeneous analysis of the FMR over all redshifts is required to provide a conclusive answer.

We studied the location of our local and high-redshift galaxies on the mass-metallicity relation and fundamental metallicity relation in Section \ref{sec:mzr}. We only use $T_e$-metallicities and therefore avoid some of the potential issues associated with strong-line calibrations. Our results in Figures \ref{fig:fmr} and \ref{fig:delta_fmr} show that for the galaxies in our samples and collated literature there may be a moderate deviation from the $z=0$ FMR at higher redshifts. This deviation is most strongly present when using the \cite{curti2020} parametrisation, however, here we also see a small systematic offset towards lower metallicities with our local DESI EDR [\textsc{Oiii}]$\lambda$4363 sample. This systematic offset is similar to the average offset observed for local blueberry and green pea galaxies \citep{yang2017}. We find that regardless of which FMR parametrisation is used, there are some indications of possible evolution where galaxies at high-redshift are more offset from the local FMR. However, this trend has very low significance ($\sim1.3\sigma$). Larger samples of galaxies may improve these constraints. More significant deviations from the FMR are found on an individual galaxy basis, several galaxies in our EXCELS and $z>2$ literature samples have a larger than $5\sigma$ discrepancy.

The growing sample of $T_e$-metallicity measurements will allow us to more conclusively answer whether a deviation from the local FMR is observed in the early Universe. These growing samples may also allow us to quantify any evolution of the FMR and compare whether these results are in agreement with simulations; several simulations do predict some evolution of the FMR \citep[e.g.,][]{garcia2024}. These deviations could be an exciting probe of an early era of galaxy evolution where a balance between gas accretion, chemical evolution and feedback processes has not yet been reached \citep{tacconi2020review}.

\section{Conclusions} \label{sec:conclusions}
We have presented $T_e$-metallicity measurements of 22 galaxies at $1.65<z<7.92$ with a median redshift $\langle z \rangle=4.05$ which were observed as part of the \textit{JWST}/EXCELS survey. We also analysed a high-quality sample of 782 local galaxies which were observed as part of the DESI Early Data Release. Each of these galaxies were selected to have a detection of the [\textsc{Oiii}]$\lambda$4363 auroral emission line. We derived key physical properties for the galaxies in these samples, such as stellar mass, star formation rate, emission line fluxes, electron temperatures, oxygen abundances and nitrogen abundances. We used these measurements for two main purposes: (1) the assessment of strong-line calibrations both in the local Universe and at high-redshift, and (2) to study the evolution of the mass-metallicity relation and fundamental metallicity relation.

Our key findings on strong-line calibrations are:
\begin{itemize}
    \item Due to a strong dependence on the ionisation parameter many strong-line calibrations (e.g., $R2$, $R3$, $O32$, $O3N2$, $Ne3O2$ and $S2$) can only be used to derive reliable metallicities when this secondary dependence is taken into account. This can be achieved through e.g., explicitly parametrising the ionisation parameter as a second parameter in the calibration or constructing a calibration sample with the same ionisation parameter range as the inference sample.
    \item The recently defined $\widehat{R}$ diagnostic by  \cite{laseter2024} provides a reliable metallicity diagnostic with minimal scatter. As this projection removes the secondary dependency on ionisation parameter it is able to provide metallicity measurements for both local and high-redshift galaxies \citep[which are systematically biased to high ionisation parameter, e.g.,][]{steidel2014, shapley2015, shapley2024}. We do find a slightly shallower calibration than \cite{laseter2024}, which is presented in Equation \ref{eq:rhat_this_work} and Figure \ref{fig:Rhat_calibration}. Further work is needed to fully understand the nature of the differences between the relation found here and the original version.
    \item We introduce the $\widehat{RNe}$ calibration (Eq. \ref{eq:nehat_diagnostic} and \ref{eq:nehat_this_work}) in Section \ref{sec:neon_sulfur_calibrators} which is an ionisation parameter independent metallicity calibration that can be used to determine the metallicity of galaxies up to very high-redshift ($z\sim11.2$ with \textit{JWST}/NIRSpec). This line diagnostic is also insensitive to dust attenuation due to the close proximity of all the required emission lines. This line ratio is particularly useful in the regime between $9.5<z<11.2$ where $\widehat{R}$ can no longer be used.
    \item The $\widehat{R}$ and $\widehat{RNe}$ are promising strong-line calibrations, however, there is still a region at $\widehat{R}\gtrsim0.7$ where these diagnostics are largely insensitive to metallicities between $7.8 \lesssim 12+\rm log(O/H) \lesssim 8.3$; due to this there is an increased uncertainty on metallicities derived using these diagnostics in this range.
    \item Emission line calibrations based on the [\textsc{Nii}]$\lambda$6584 emission line may be biased at high-redshift due to systematic evolution of the N/O abundance ratio. We find that many of the EXCELS galaxies have moderately enhanced $\rm log(N/O)$ abundance ratios at fixed $\rm log(O/H)$ compared to local galaxies in our DESI EDR sample. This issue does not affect line diagnostics using neon or sulfur emission lines as these are like oxygen also alpha-elements \citep[e.g.,][]{henry1989, garnett2002}.
    \item Non-parametric metallicity calibrations such as those based on a KDE derived from a calibration sample \citep[similar to e.g.,][]{langeroodi2024} provide competitive results to traditional parametric calibrations.
\end{itemize}

Our findings on the evolution of the MZR and FMR are:
\begin{itemize}
    \item We find that the EXCELS galaxies in our sample are located offset below the $z=0$ MZR at very similar stellar mass and metallicity to other [\textsc{Oiii}]$\lambda$4363 detected galaxies at $z>2$ in recent \textit{JWST} and ground based observations \citep[][]{sanders2023metallicity,curti2023,nakajima2023,morishita2024, cullen2025}. We also see that the at fixed stellar mass the metallicities of the EXCELS galaxies decrease with redshift.
    \item There may be a mild systematic evolution of the FMR as a function of redshift at $z>4$ \citep[as also shown by][]{curti2024}. However, similar to \cite{nakajima2023} we find that the strength of this evolution is dependent on the FMR parametrisation used. The significance of this systematic offset from the FMR is low (a maximum of $\sim1.3\sigma$ for the \cite{curti2020} parametrisation) on a population basis due to the small sample of $T_e$-measurements in our work.
    \item On an individual basis we find several galaxies in our EXCELS and literature samples which are significantly offset from the local FMR at larger than $5\sigma$ significance. These galaxies, which we find particularly at $z>3$, may be a signature of an era in galaxy evolution where the equilibrium between gas accretion, chemical evolution and feedback processes has not yet been reached. However, this does not fully account for all the systematic uncertainties involved in the measurements. 
    \item The reliable measurement of a redshift evolution of the FMR requires a homogeneous analysis of local and high-redshift galaxies with a large sample of $T_e$-metallicity measurements (for which the current sample size is still limited) or unbiased strong-line calibrated metallicities. 
\end{itemize}

In this work we highlight the fast growing number of galaxies for which we can reliably measure chemical abundances using the $T_e$-method. This is shown by the large number of high-redshift galaxies where deep \textit{JWST} spectroscopy allows us to measure faint auroral emission lines in galaxy spectra, in our case using the \textit{JWST} EXCELS survey. We also show the rapid growth of local galaxies with $T_e$-abundances. Over the next few years the DESI survey will increase the number of local $T_e$-abundance measurements of galaxies by an order of magnitude. These growing observational capabilities at all redshifts will allow us to make significant improvements to strong-line metallicity calibrations which are applicable at all redshifts. These constraints on chemical evolution will also be an important probe of the baryon cycle in the early Universe. We will be able to probe the chemical evolution of galaxies in exquisite detail with the combined samples that are being acquired by DESI, \textit{JWST} and soon MOONS, PFS, WEAVE and 4MOST.

\section*{Acknowledgements}
% Add acknowledgements
D. Scholte, F. Cullen, K. Z. Arellano-C\'ordova and T. M. Stanton and acknowledge support from a UKRI Frontier Research Guarantee Grant (PI Cullen; grant reference EP/X021025/1).
A. C. Carnall, H.-H. Leung and S. Stevenson acknowledge support from a UKRI Frontier Research Guarantee Grant (PI Carnall; grant reference EP/Y037065/1). 
J. S. Dunlop thanks the Royal Society for support via a Research Professorship.
H. Zou acknowledges the supports from National Key R\&D Program of China (grant Nos. 2023YFA1607804 and 2022YFA1602902), and the National Natural Science Foundation of China (NSFC; grant Nos. 12120101003, 12373010).

% Mandatory DESI EDR acknowledgement
This research used data obtained with the Dark Energy Spectroscopic Instrument (DESI). DESI construction and operations is managed by the Lawrence Berkeley National Laboratory. This material is based upon work supported by the U.S. Department of Energy, Office of Science, Office of High-Energy Physics, under Contract No. DE–AC02–05CH11231, and by the National Energy Research Scientific Computing Center, a DOE Office of Science User Facility under the same contract. Additional support for DESI was provided by the U.S. National Science Foundation (NSF), Division of Astronomical Sciences under Contract No. AST-0950945 to the NSF’s National Optical-Infrared Astronomy Research Laboratory; the Science and Technology Facilities Council of the United Kingdom; the Gordon and Betty Moore Foundation; the Heising-Simons Foundation; the French Alternative Energies and Atomic Energy Commission (CEA); the National Council of Science and Technology of Mexico (CONACYT); the Ministry of Science and Innovation of Spain (MICINN), and by the DESI Member Institutions: www.desi.lbl.gov/collaborating-institutions. The DESI collaboration is honored to be permitted to conduct scientific research on Iolkam Du’ag (Kitt Peak), a mountain with particular significance to the Tohono O’odham Nation. Any opinions, findings, and conclusions or recommendations expressed in this material are those of the author(s) and do not necessarily reflect the views of the U.S. National Science Foundation, the U.S. Department of Energy, or any of the listed funding agencies.

\section*{Data Availability}
\label{sec:data_availability}
All \textit{JWST} and HST data products are available via the Mikulski Archive for Space Telescopes (\url{https://mast.stsci.edu}). 
All the DESI data products used here are available as part of the DESI Early Data Release (\url{https://data.desi.lbl.gov/doc/}). The derived data products presented in Tables \ref{tab:excels_derived_data}, \ref{tab:desi_derived_data} and \ref{tab:line_ratios} will be made available in machine readable format (\url{https://dirkscholte.github.io/data}).
Additional data products are available from the authors upon reasonable request.

\bibliography{main}

\appendix

\section{Residuals of strong-line diagnostics}
\label{sec:appendix_residuals_diagnostics}
We assess the performance of strong-line calibrations through their residuals on our EXCELS and DESI EDR samples. We use two different methods to calculate the residuals: (1) the $\Delta R = R-R_{\rm cal}(Z_{T_e})$ and (2) the $\Delta \log({\rm O/H}) = Z_{\rm cal}(R) - Z_{T_e}$. The first residual measurement ($\Delta R$) is useful to describe the dispersion in the fitted strong-line calibration. The second residual measurement ($\Delta \log(\rm O/H)$) gives an indication of the bias and offset of derived strong-line metallicities using a particular calibration.

\begin{figure*}
	\begin{center}
	\includegraphics[width=0.95\textwidth]{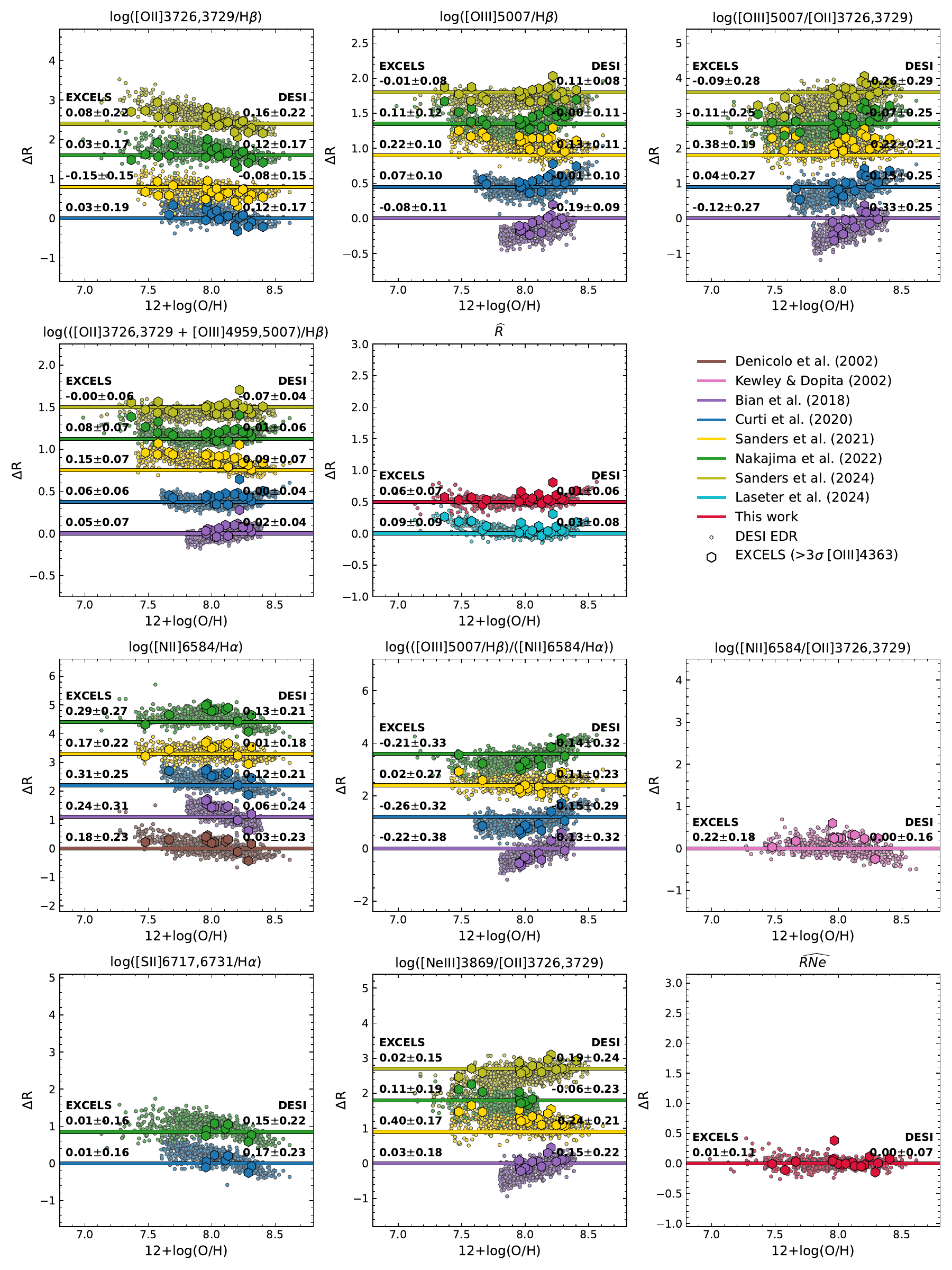}
	\caption{The line ratio residuals ($\Delta R$) between measured emission line ratis and the line ratios expected for each calibration at the $T_e$-metallicity for EXCELS (\textit{hexagons, black edges}) and DESI EDR galaxies (\textit{circles, black edges}). The \textit{R2}, \textit{R3}, \textit{O32}, \textit{R23}, $\widehat{R}$, \textit{N2}, \textit{O3N2}, \textit{N2O2}, \textit{S2}, \textit{Ne3O2} and $\widehat{RNe}$ (\textit{left-to-right} and \textit{top-to-bottom}) strong-line ratios are shown in the individual panels. The residual data points are coloured according to the strong-line calibration used as shown in the legend. The residuals of each strong-line calibration are offset. The offset is scaled by the dynamic range of the measured line ratios of each diagnostic.}
	\label{fig:auroral_lineratios_residuals}
	\end{center}
\end{figure*}

In Figure \ref{fig:auroral_lineratios_residuals} we show the $\Delta R$ residuals calculated for the different strong-line calibrations. The average stated residuals for each calibration are stated for the EXCELS and DESI EDR samples in the form of bias$\pm$scatter. We only include residuals of galaxies with direct metallicity measurements within the stated range of each of the strong-line diagnostics. If no metallicity range is specified we use the range of the calibration data used.

\begin{figure*}
	\begin{center}
	\includegraphics[width=0.95\textwidth]{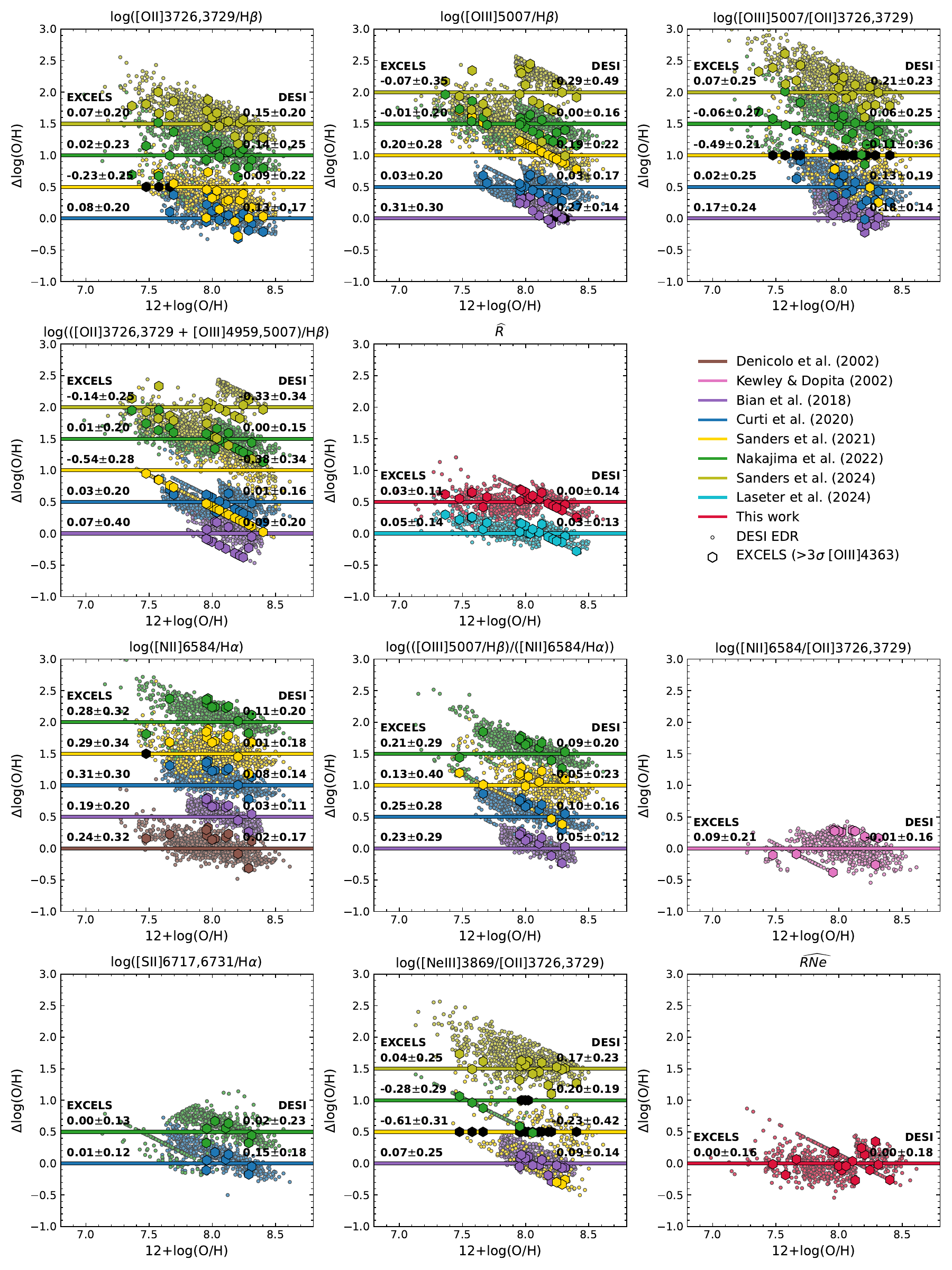}
	\caption{The metallicity residuals ($\Delta \log(\rm O/H)$) between $T_e$-metallicities and strong-line diagnostics for EXCELS (\textit{hexagons, black edges}) and DESI EDR galaxies (\textit{circles, black edges}). The \textit{R2}, \textit{R3}, \textit{O32}, \textit{R23}, $\widehat{R}$, \textit{N2}, \textit{O3N2}, \textit{N2O2}, \textit{S2}, \textit{Ne3O2} and $\widehat{RNe}$ (\textit{left-to-right} and \textit{top-to-bottom}) strong-line ratios are shown in the individual panels. The residual data points are coloured according to the strong-line calibration used as shown in the legend. The residuals of each strong-line calibration are offset by 0.5 dex.}
	\label{fig:auroral_lineratios_metallicity_residuals}
	\end{center}
\end{figure*}

In Figure \ref{fig:auroral_lineratios_metallicity_residuals}, where we look at the residuals between the strong-line metallicity and direct metallicity ($\Delta \log({\rm O/H}) = \log({\rm O/H})_{\rm strong} - \log({\rm O/H})_{T_e}$). In each panel we show the residuals compared to each of the diagnostics for both EXCELS and DESI EDR galaxies. We also note the bias and 1$\sigma$ scatter for these samples on the left and right side, respectively. There are several important choices we made in the determination of the residual measurements:
\begin{itemize}
\item We only include residuals of galaxies with direct metallicity measurements within the stated range of each of the strong-line diagnostics. If no metallicity range is specified we use the range of the calibration data used.
\item We exclude residuals galaxies with strong-line ratios which are suggestive of metallicities above or below the metallicity range of the diagnostic (these data points are shown in black on the zero line of the calibration). 
\item For non-monotonic metallicity diagnostics we may derive multiple metallicity solutions for a strong-line ratio, in this case we only include the inferred metallicity closest to the $T_e$-metallicity measurement. 
\item We do include the inferred metallicity measurement of non-monotonic diagnostics where the strong-line ratio is beyond the turn-over ratio of the diagnostic; e.g., a galaxy with a $R23$ value of 1.1 will be included in the residual measurements of the \cite{nakajima2022} $R23$ diagnostic. We include these residual measurements as this is a common approach used when inferring metallicities using strong-line diagnostics.
\end{itemize}
These choices resemble the best case scenario for the performance of the strong-line diagnostics. One further caveat to the interpretation of these residuals is that the large differences in the metallicity ranges covered by different diagnostics need to be taken into account when comparing the residuals of multiple diagnostics; a diagnostic with a small metallicity range will typically have a smaller residual scatter. In the remainder of this section we will discuss some of these strong-line diagnostics individually.

\section{An alternative $\widehat{R}$-projection} \label{sec:rhat_chakraborty}

\begin{figure}
	\begin{center}
	\includegraphics[width=0.495\textwidth]{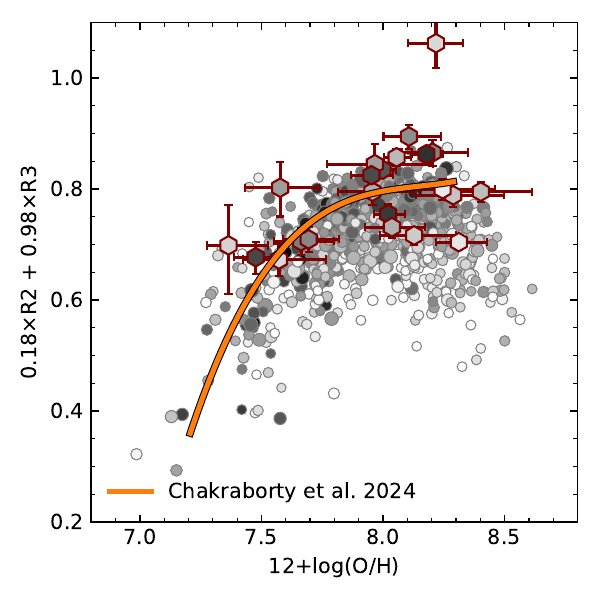}
	\caption{The galaxies in our EXCELS and DESI samples shown on the $\widehat{R}$ reprojection proposed by \protect\cite{chakraborty2024}. The details of the Figure are described by the legend in Figure \ref{fig:auroral_lineratios_hb_ew_oxygen}.}
	\label{fig:rhat_chakraborty}
	\end{center}
\end{figure}

A recent paper by \cite{chakraborty2024} proposed a reprojection of the $\widehat{R}$-diagnostic introduced by \cite{laseter2024}. Based on a calibration sample of high-redshift galaxies they showed that a more narrow sequence than the \cite{laseter2024} $\widehat{R}$ can be achieved by using a different relative combination of $R2$ and $R3$: $\widehat{R}_{\rm Chakraborty} = 0.18\times R2 + 0.98 \times R3$. We show our EXCELS and DESI galaxies in this reprojection in Figure \ref{fig:rhat_chakraborty}. As is visible through the comparison of Figures \ref{fig:rhat_chakraborty} and \ref{fig:Rhat_calibration}, the projection proposed by \cite{chakraborty2024} does not provide an improvement over the \cite{laseter2024} projection for our EXCELS and DESI samples. This new projection shows a dependency on ionisation parameter (as measured through EW(H$\beta$)). It is, however, important that with increasingly large samples we revisit this question to determine the optimal $\widehat{R}$-projection.

\section{EXCELS emission line diagnostics}
The logarithmic values for the emission line diagnostics used in this work are listed in Table \ref{tab:line_ratios}.
\begin{landscape}
\renewcommand{\arraystretch}{1.25}
\begin{table}
\centering
\caption{The measured line ratios for the EXCELS sample. All line ratios are in logarithmic form. Values denoted with a dagger$^{\dagger}$ are 2-$\sigma$ upper- or lower-limits.}
\begin{tabular}{cccccccccccc}

\hline 
Target ID & \textit{R2} & \textit{R3} & \textit{O32} & \textit{R23} & $\widehat{R}$ & \textit{N2} & \textit{O3N2} & \textit{N2O2} & \textit{S2} & \textit{Ne3O2} & $\widehat{RNe}$ \\
\hline \hline

40081 & ${0.210}_{-0.040}^{+0.037}$ & ${0.731}_{-0.011}^{+0.010}$ & ${0.520}_{-0.045}^{+0.041}$ & ${0.937}_{-0.009}^{+0.008}$ & ${0.742}_{-0.028}^{+0.026}$ & ${-1.293}_{-0.019}^{+0.018}$ & ${2.025}_{-0.054}^{+0.048}$ & ${-1.069}_{-0.100}^{+0.081}$ & ${-0.998}_{-0.013}^{+0.013}$ & ${-0.523}_{-0.016}^{+0.016}$ & ${0.216}_{-0.033}^{+0.030}$ \\ 
45177 & ${0.214}_{-0.056}^{+0.050}$ & ${0.873}_{-0.012}^{+0.012}$ & ${0.659}_{-0.055}^{+0.049}$ & ${1.064}_{-0.053}^{+0.047}$ & ${0.869}_{-0.037}^{+0.034}$ & ${-1.147}_{-0.032}^{+0.030}$ & ${2.020}_{-0.005}^{+0.005}$ & ${-0.914}_{-0.069}^{+0.059}$ & ${-1.038}_{-0.027}^{+0.025}$ & ${-0.441}_{-0.019}^{+0.018}$ & ${0.261}_{-0.038}^{+0.035}$ \\ 
45393 & ${0.288}_{-0.042}^{+0.038}$ & ${0.693}_{-0.011}^{+0.011}$ & ${0.404}_{-0.045}^{+0.041}$ & ${0.935}_{-0.012}^{+0.011}$ & ${0.745}_{-0.030}^{+0.028}$ & --- & --- & --- & --- & --- & --- \\ 
47557 & ${0.241}_{-0.063}^{+0.055}$ & ${0.766}_{-0.013}^{+0.013}$ & ${0.525}_{-0.061}^{+0.053}$ & ${0.978}_{-0.054}^{+0.048}$ & ${0.788}_{-0.041}^{+0.037}$ & ${-1.307}_{-0.049}^{+0.044}$ & ${2.074}_{-0.005}^{+0.005}$ & ${-1.103}_{-0.085}^{+0.071}$ & ${-1.238}_{-0.058}^{+0.051}$ & ${-0.381}_{-0.045}^{+0.041}$ & ${0.299}_{-0.096}^{+0.079}$ \\ 
48659 & ${-0.309}_{-0.258}^{+0.160}$ & ${0.769}_{-0.046}^{+0.042}$ & ${1.078}_{-0.254}^{+0.159}$ & ${0.922}_{-0.047}^{+0.042}$ & ${0.531}_{-0.157}^{+0.115}$ & $<{-0.831}$$^{\dagger}$ & $>{1.601}$$^{\dagger}$ & $<{-0.071}$$^{\dagger}$ & --- & ${-0.278}_{-0.426}^{+0.211}$ & ${-0.180}_{-0.486}^{+0.223}$ \\ 
52422 & ${-0.160}_{-0.091}^{+0.075}$ & ${0.720}_{-0.014}^{+0.014}$ & ${0.880}_{-0.095}^{+0.077}$ & ${0.884}_{-0.054}^{+0.048}$ & ${0.558}_{-0.054}^{+0.048}$ & ${-2.035}_{-0.267}^{+0.164}$ & ${2.755}_{-0.059}^{+0.052}$ & ${-1.439}_{-0.330}^{+0.185}$ & --- & ${-0.097}_{-0.077}^{+0.065}$ & ${0.015}_{-0.079}^{+0.066}$ \\ 
56875 & ${0.147}_{-0.035}^{+0.032}$ & ${0.659}_{-0.022}^{+0.021}$ & ${0.512}_{-0.030}^{+0.028}$ & ${0.879}_{-0.022}^{+0.021}$ & ${0.649}_{-0.037}^{+0.034}$ & $<{-1.213}$$^{\dagger}$ & $>{1.873}$$^{\dagger}$ & $<{-1.067}$$^{\dagger}$ & --- & ${-0.680}_{-0.083}^{+0.070}$ & ${0.094}_{-0.114}^{+0.090}$ \\ 
57498 & ${0.173}_{-0.016}^{+0.015}$ & ${0.820}_{-0.007}^{+0.007}$ & ${0.646}_{-0.015}^{+0.015}$ & ${1.017}_{-0.007}^{+0.007}$ & ${0.802}_{-0.014}^{+0.013}$ & ${-1.347}_{-0.029}^{+0.027}$ & ${2.167}_{-0.003}^{+0.003}$ & ${-1.075}_{-0.036}^{+0.033}$ & --- & ${-0.443}_{-0.021}^{+0.020}$ & ${0.226}_{-0.028}^{+0.026}$ \\ 
59009 & ${0.043}_{-0.021}^{+0.020}$ & ${0.872}_{-0.009}^{+0.009}$ & ${0.828}_{-0.021}^{+0.020}$ & ${1.044}_{-0.009}^{+0.009}$ & ${0.788}_{-0.018}^{+0.017}$ & --- & --- & --- & --- & ${-0.303}_{-0.021}^{+0.020}$ & ${0.194}_{-0.032}^{+0.029}$ \\ 
59720 & ${0.245}_{-0.026}^{+0.025}$ & ${0.685}_{-0.011}^{+0.011}$ & ${0.440}_{-0.026}^{+0.025}$ & ${0.929}_{-0.011}^{+0.011}$ & ${0.718}_{-0.022}^{+0.021}$ & ${-1.042}_{-0.039}^{+0.036}$ & ${1.728}_{-0.052}^{+0.047}$ & ${-0.904}_{-0.080}^{+0.068}$ & ${-0.889}_{-0.037}^{+0.034}$ & ${-0.611}_{-0.030}^{+0.028}$ & ${0.202}_{-0.045}^{+0.041}$ \\ 
63962 & ${-0.151}_{-0.044}^{+0.040}$ & ${0.747}_{-0.017}^{+0.016}$ & ${0.898}_{-0.041}^{+0.038}$ & ${0.931}_{-0.017}^{+0.016}$ & ${0.587}_{-0.036}^{+0.033}$ & ${-1.688}_{-0.104}^{+0.083}$ & ${2.436}_{-0.052}^{+0.046}$ & ${-1.300}_{-0.129}^{+0.099}$ & --- & ${-0.164}_{-0.056}^{+0.050}$ & ${0.154}_{-0.080}^{+0.067}$ \\ 
69991 & ${0.437}_{-0.082}^{+0.069}$ & ${0.781}_{-0.025}^{+0.023}$ & ${0.344}_{-0.085}^{+0.071}$ & ${1.037}_{-0.027}^{+0.025}$ & ${0.893}_{-0.060}^{+0.053}$ & ${-1.125}_{-0.119}^{+0.093}$ & ${1.906}_{-0.052}^{+0.047}$ & ${-1.112}_{-0.217}^{+0.144}$ & --- & ${-0.480}_{-0.062}^{+0.054}$ & ${0.609}_{-0.111}^{+0.088}$ \\ 
70864 & ${-0.052}_{-0.039}^{+0.036}$ & ${0.892}_{-0.016}^{+0.016}$ & ${0.945}_{-0.039}^{+0.035}$ & ${1.054}_{-0.016}^{+0.015}$ & ${0.761}_{-0.033}^{+0.031}$ & ${-1.372}_{-0.098}^{+0.080}$ & ${2.265}_{-0.054}^{+0.048}$ & ${-0.918}_{-0.132}^{+0.101}$ & --- & ${-0.190}_{-0.036}^{+0.033}$ & ${0.220}_{-0.059}^{+0.052}$ \\ 
73535 & ${0.257}_{-0.033}^{+0.031}$ & ${0.764}_{-0.011}^{+0.011}$ & ${0.506}_{-0.035}^{+0.033}$ & ${0.982}_{-0.011}^{+0.011}$ & ${0.793}_{-0.026}^{+0.024}$ & --- & --- & --- & --- & ${-0.557}_{-0.022}^{+0.021}$ & ${0.279}_{-0.040}^{+0.037}$ \\ 
93897 & ${0.310}_{-0.032}^{+0.030}$ & ${1.027}_{-0.038}^{+0.035}$ & ${0.717}_{-0.032}^{+0.030}$ & ${1.211}_{-0.038}^{+0.035}$ & ${1.049}_{-0.050}^{+0.044}$ & --- & --- & --- & --- & --- & --- \\ 
94335 & ${0.294}_{-0.010}^{+0.010}$ & ${0.664}_{-0.005}^{+0.005}$ & ${0.370}_{-0.010}^{+0.009}$ & ${0.917}_{-0.005}^{+0.005}$ & ${0.722}_{-0.010}^{+0.009}$ & ${-0.947}_{-0.011}^{+0.011}$ & ${1.611}_{-0.001}^{+0.001}$ & ${-0.788}_{-0.018}^{+0.018}$ & ${-0.902}_{-0.055}^{+0.049}$ & ${-0.701}_{-0.016}^{+0.015}$ & ${0.231}_{-0.022}^{+0.021}$ \\ 
95839 & ${0.085}_{-0.032}^{+0.030}$ & ${0.858}_{-0.010}^{+0.010}$ & ${0.773}_{-0.035}^{+0.033}$ & ${1.035}_{-0.009}^{+0.009}$ & ${0.796}_{-0.024}^{+0.023}$ & --- & --- & --- & --- & ${-0.358}_{-0.018}^{+0.018}$ & ${0.236}_{-0.034}^{+0.032}$ \\ 
104937 & ${0.397}_{-0.034}^{+0.032}$ & ${0.742}_{-0.013}^{+0.013}$ & ${0.345}_{-0.035}^{+0.033}$ & ${0.998}_{-0.014}^{+0.013}$ & ${0.840}_{-0.028}^{+0.026}$ & --- & --- & --- & --- & ${-0.657}_{-0.037}^{+0.034}$ & ${0.354}_{-0.055}^{+0.048}$ \\ 
119504 & ${-0.207}_{-0.109}^{+0.087}$ & ${0.856}_{-0.033}^{+0.030}$ & ${1.063}_{-0.107}^{+0.086}$ & ${1.012}_{-0.033}^{+0.030}$ & ${0.656}_{-0.080}^{+0.068}$ & --- & --- & --- & --- & ${0.044}_{-0.104}^{+0.084}$ & ${-0.029}_{-0.117}^{+0.092}$ \\ 
121806 & ${-0.040}_{-0.033}^{+0.031}$ & ${0.848}_{-0.013}^{+0.012}$ & ${0.889}_{-0.034}^{+0.031}$ & ${1.014}_{-0.012}^{+0.012}$ & ${0.728}_{-0.027}^{+0.025}$ & ${-1.196}_{-0.048}^{+0.044}$ & ${2.045}_{-0.053}^{+0.047}$ & ${-0.762}_{-0.093}^{+0.076}$ & ${-1.399}_{-0.120}^{+0.094}$ & ${-0.158}_{-0.027}^{+0.025}$ & ${0.266}_{-0.045}^{+0.041}$ \\ 
123597 & ${0.154}_{-0.039}^{+0.035}$ & ${0.776}_{-0.015}^{+0.014}$ & ${0.622}_{-0.040}^{+0.036}$ & ${0.973}_{-0.014}^{+0.014}$ & ${0.755}_{-0.031}^{+0.029}$ & ${-1.570}_{-0.109}^{+0.087}$ & ${2.347}_{-0.054}^{+0.048}$ & ${-1.307}_{-0.144}^{+0.108}$ & ${-1.114}_{-0.048}^{+0.043}$ & ${-0.661}_{-0.050}^{+0.045}$ & ${0.083}_{-0.069}^{+0.059}$ \\ 
123837 & ${0.120}_{-0.037}^{+0.034}$ & ${0.701}_{-0.016}^{+0.015}$ & ${0.581}_{-0.038}^{+0.035}$ & ${0.915}_{-0.015}^{+0.015}$ & ${0.673}_{-0.032}^{+0.029}$ & --- & --- & --- & --- & --- & --- \\ 
\hline
\end{tabular}
\label{tab:line_ratios}
\end{table}
\renewcommand{\arraystretch}{1.}

\label{lastpage}
\end{landscape}
\end{document}